# Quantitative kinetic rules for plastic strain-induced α - ω phase transformation in Zr under high pressure

Achyut Dhar[1,5*], Valery I. Levitas[1,2*], K. K. Pandey[1,3,4], Changyong Park[5], Maddury Somayazulu[5], Nenad Velisavljevic[5,6]

[1]Department of Aerospace Engineering, Iowa State University, Ames, IA 50011, USA
[2]Department of Mechanical Engineering, Iowa State University, Ames, IA 50011, USA
[3]High Pressure & Synchrotron Radiation Physics Division, Bhabha Atomic Research Centre, Bombay, Mumbai-400085, India (present affiliation)
[4]Homi Bhabha National Institute, Anushaktinagar, Mumbai 400094, India (present affiliation)
[5]HPCAT, X-ray Science Division, Argonne National Laboratory, Argonne, Illinois 60439, USA
[6]Lawrence Livermore National Laboratory, Physics Division, Livermore, CA 94550, USA

* Corresponding authors: Email: vlevitas@iastate.edu and adhar@iastate.edu

These authors contributed equally: Valery I. Levitas, Achyut Dhar

## ABSTRACT

Plastic strain-induced phase transformations (PTs) and chemical reactions under high pressure are broadly spread in modern technologies[2,28,29], friction and wear[3,12,13], geophysics[1,3,15,19], and astrogeology[20,21]. However, because of very heterogeneous fields of plastic strain $\boldsymbol{E}^p$ and stress $\boldsymbol{\sigma}$ tensors and volume fraction $c$ of phases in a sample compressed in a diamond anvil cell (DAC) and impossibility of measurements of $\boldsymbol{\sigma}$ and $\boldsymbol{E}^p$, there are no strict kinetic equations for them. Here, we develop combined experimental-computational approaches to determine all fields in strongly plastically predeformed Zr and kinetic equation for α-ω PT consistent with experimental data for the entire sample. Kinetic equation depends on accumulated plastic strain (instead of time) and pressure and is independent of plastic strain and deviatoric stress tensors, i.e., it can be applied for various above processes. Our results initiate kinetic studies of strain-induced PTs and provide efforts toward more comprehensive understanding of material behavior in extreme conditions.

In comparison with hydrostatic loading, plastic straining drastically decreases pressure for PT[14,15,18] (and chemical reactions[6,20,21], which are not central part of the current work), produces new phases, alters PT kinetics from time-dependent to plastic strain dependent, replaces reversible PTs with irreversible, and produces nanostructured materials[7,14,15,17,18,28]. That leads these PTs into a special category, strain-induced PTs[16,26,27]. In contrast to traditional pressure-induced PTs, which originate at pre-existing defects that cause stress concentration, strain-induced PTs initiate at defects generated during plastic flow. The only existing defect that can reduce PT pressure by one to two orders of magnitude or up to 70 GPa (observed for graphite-diamond PT[15]) is a dislocation pileup, as confirmed by analytical[16], atomistic[4], and phase field[5,9] modeling. While there are successful in-situ studies of strain-induced PTs under compression in DAC (Fig. 1a)[8,10,18,27,28] and torsion in rotational DAC, there is one major challenge: *no strict kinetic equations for strain-induced PTs*. The reason for the lack of fundamental equations is that since they occur under the action of plastic strain $\boldsymbol{E}^p$ and stress $\boldsymbol{\sigma}$ tensors, the corresponding kinetics should depend on $\boldsymbol{E}^p$ and $\boldsymbol{\sigma}$; however, neither $\boldsymbol{E}^p$ nor $\boldsymbol{\sigma}$ are experimentally measurable. There are only simplified empirical expressions. For example, in ball milling, kinetics of trimerization reaction is determined in terms of number of balls' impacts[2]. During high-pressure torsion, the volume fraction $c$ of ω-Zr during α-ω PT was determined as a function of shear strain (defined by a simplified equation) for one of the radii[7], in which $c$ was measured postmortem after unloading. Moreover, the effect of pressure $p$ was not included, and it was not verified whether the proposed kinetics is valid for other radii. In recent in-situ measurements for compression in DAC and torsion in rotational DAC, the kinetic equation $dc/dq=f(q,p,c)$ for α-ω PT in Zr (where $q$ is the accumulated plastic strain) was based on data at the sample symmetry axis only[10]. While radial $\bar{c}$ and $\bar{p}$ distributions were measured as an average over the sample thickness, the averaged $\bar{q}$ could only be approximately evaluated at the sample symmetry axis. There, one can assume homogenous $c$, $p$, and $q$ fields along the sample thickness, implying that material is under uniaxial compression. However, such equation was not checked for finite radii, where significant and heterogeneous shears are present, because $q$ cannot be strictly evaluated without simulations. Also, as we will see in Fig. 4, $c$ and $q$ are strongly heterogeneous along the symmetry axis too. Another global problem is that all constitutive (including kinetics) equations should be determined for material points (i.e., Lagrangian view of the motion). However, the X-ray measurements are performed in spatial points (i.e., Eulerian description), and due to large plastic flow, different material points pass through the X-ray beam at different loads. Independently, theoretically derived kinetics[16] was used for finite element analysis (FEA) of the processes in traditional[22,24] and rotational[23,25] DAC but with some model parameters.

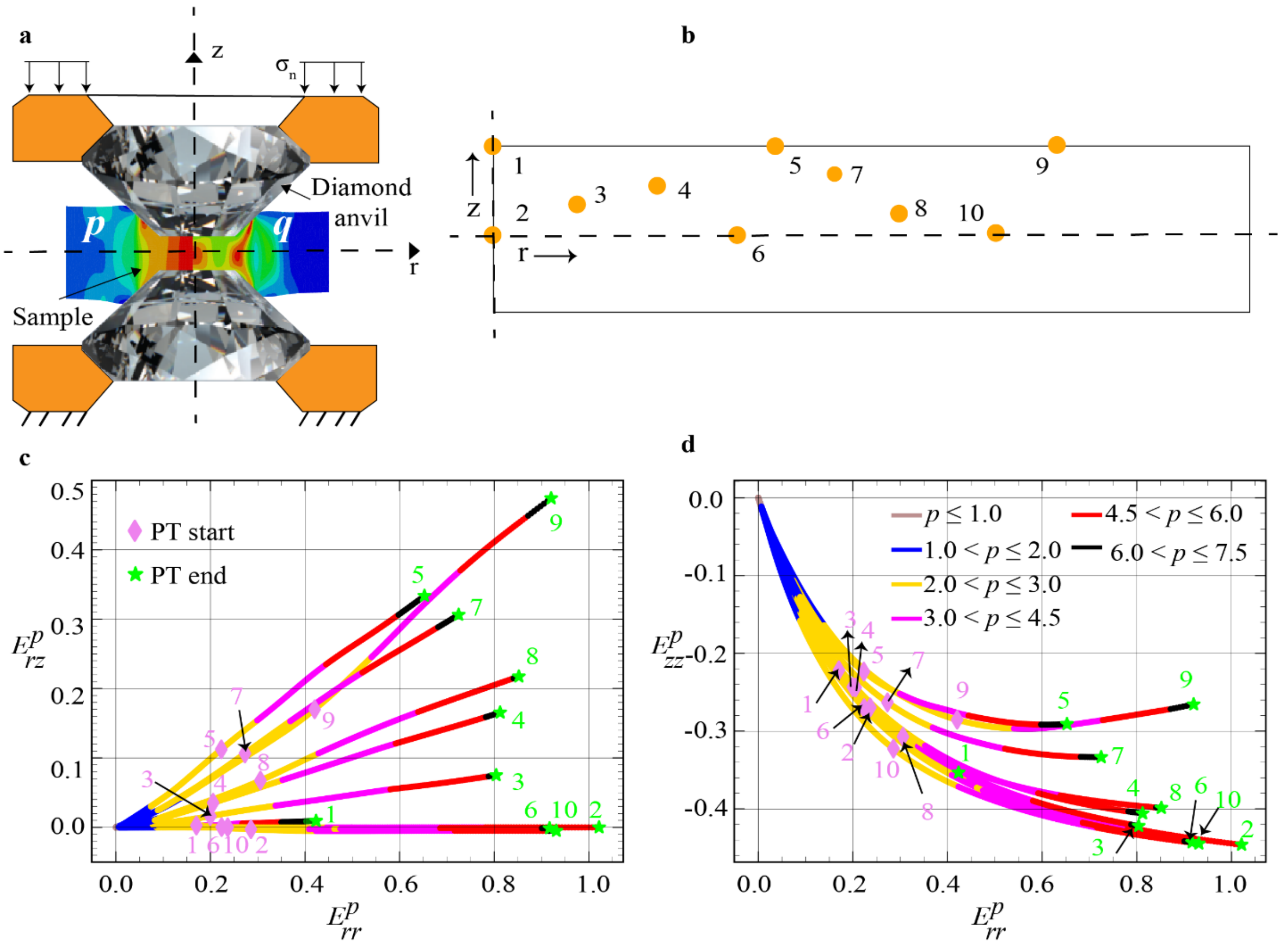

**Fig. 1 2D projections of the 4D plastic strain – pressure trajectories. a** DAC schematic. **b** Cross-section of the sample showing the positions of the material points in the undeformed configuration. The coordinates of the material points are: #1(0, 82.5); #2(0, 0); #3(11.3, 42.1); #4(29.2, 45.5); #5(54.1, 82.5); #6(37.8, 0.0); #7(52.3, 66.5); #8(57.2, 40.2); #9(93, 82.5); #10(76.2, 0.0). The units are in $\mu m$. **c** and **d** Shear $E^p_{rz}$ vs. radial $E^p_{rr}$ and axial $E^p_{zz}$ vs. radial $E^p_{rr}$ plastic strain trajectories, respectively, for various material points with superposed pressure $p$ evolution (in GPa) with parameters $\delta_1 = \delta_3 = 0, \delta_2 = 0.803, k = 5.20$, and $p^d_\varepsilon = 2.65 - (q_0 - 0.42)$ in the kinetic equation.

Generally, kinetic equation for strain-induced PTs should have a form $dc/dq=f(\boldsymbol{E}^p,\boldsymbol{\sigma},c)$. Recently, we developed coupled experimental-analytical-computational approaches for finding fields of $\boldsymbol{E}^p$ and $\boldsymbol{\sigma}$ in the entire DAC Zr sample[8]. Part of this method, coupled experimental-analytical (CEA) approach, was utilized for determining the distribution of the friction stress between sample and diamond. PT was not modeled, and the field of $\bar{c}$ averaged over the sample thickness from experiments was used as input data uniformly along the thickness. Here, we develop a new Combined Experimental-FEM approach (CE-FEM), which includes modeling α-ω PT and determining $\boldsymbol{E}^p$, $\boldsymbol{\sigma}$, and $c$ fields in a strongly plastically predeformed Zr sample and kinetic equation for α-ω PT. This is done by iteratively solving an inverse problem on determining 5 material parameters $\delta_i$ in the kinetic equation (1). Zr sample was subjected to multiple rolling until its hardness, grain size, and dislocation density no longer changed[10,11] with further straining, which exclude their effects on PT and thereby, significantly simplify the problem.

We use the same experimental data for the strongly pre-deformed Zr sample, contact friction stress distribution between diamond anvils and sample determined by CEA method from[8]. The flowchart for the developed CE-FEM approach is presented in Supplementary Fig. S1. The complete system of equations for coupled elastoplastic flow and strain-induced PTs is presented in supplementary Eqs. (S1)-(S19). FEM formulation is presented in Supplementary Notes[39]. Distribution of friction shear stress at the contact between sample and diamond obtained from[8] in addition to elastoplastic properties are used as the input data. While mechanical part is the same as in our previous large-strain models[8,24], a new kinetic equation

$$\frac{dc}{dq} = k(1+\delta_1 q)(1+\delta_3 p)(1-\delta_2 c)\frac{\sigma_{y0}^{\omega}}{c\sigma_{y0}^{\alpha}+(1-c)\sigma_{y0}^{\omega}}\left(\frac{p(q)-p_{\varepsilon}^{d}}{p_h^d-p_{\varepsilon}^{d}}\right)H(p-p_{\varepsilon}^{d}); \quad p_{\varepsilon}^{d} = \delta_4+\delta_5 q_0 \qquad (1)$$

is included and coupled to the mechanical equation, which results in a generalization in comparison with[16]. Here, $p_h^d$ is the pressure for initiation of pressure-induced PT under hydrostatic loading ($p_h^d = 5.4\ GPa$)[8], $p_{\varepsilon}^{d}$ is the minimum pressure for initiation of the plastic strain-induced PT, *p(q)* is the loading path, $\sigma_{y0}^{\alpha}$ and $\sigma_{y0}^{\omega}$are the yield strengths of the α and ω phases under ambient pressure, respectively, $q_0$ is the value of $q$ at the beginning of PT, i.e., at $p = p_{\varepsilon}^{d}$; since $q_0$=const after the PT starts, $p_{\varepsilon}^{d}$ at each material point is heterogeneously distributed constant during PT.

Eq. (1) reduces to the theoretically derived kinetics in[16], if $\delta_1 = \delta_3 = \delta_5 = 0$ and $\delta_2 = 1$, which was used in previous FEA simulations[8,22,23,24,26]. It is based on assumption that, instead of tensorial variables $\boldsymbol{E}^p$ and $\boldsymbol{\sigma}$, their scalar counterparts, $q$ and $p$, can be used. This assumption greatly simplifies theory and will be justified by experiment. Supplementary Fig. S2 shows comparison of the FEM simulated and experimental radial distributions of $\bar{c}$ for the case with $\delta_1 = \delta_3 = \delta_5 = 0$, $\delta_2 = 1$, and $p_{\varepsilon}^{d} = 2.70$ GPa. Despite the significant deviations, the general trends (i.e., the main physics) are described satisfactorily. To achieve improved quantitative correspondence, we employed two ways.

(a) Assuming $\delta_5 = 0$, i.e., constant $p_{\varepsilon}^{d}$ as claimed in[8,10], we found the simplest linear dependence of the proportionality factor on $p$ and $q$, and slightly corrected dependence on $c$, as well as slightly corrected value of $p_{\varepsilon}^{d}$. Condition $\delta_2 = 1$ implies that *dc/dq=0* for *c*=1, which is not the case for Zr. Thus, despite introducing 5 fitting parameters, obtained kinetic equation (1) remains the physics-based. After FEA of sample loading in experiments with different material parameters in Eq. (1), actual parameters have been chosen from the cumulative error *Er* minimization in Eq. (S12) between theoretical and experimental volume fractions of ω-Zr averaged over the sample thickness for all loadings and radii. The optimal set of material parameters is found as $k = 0.75$, $\delta_1 = 12.0$, $\delta_2 = 0.925$, $\delta_3 = -0.048$, and $p_{\varepsilon}^{d} = 2.65\ GPa$ with *Er*=1.19. Figs. 2a,b show comparison of the calculated and experimental radial distributions of $\bar{c}$ averaged over the sample thickness for 12 different loads. Fig. 3 presents a comparison between experiments and FEA for $\bar{c}$ versus $\bar{p}$ for different radii. Correspondence between calculated and experimental curves is remarkably good for all cases as the difference $\overline{\Delta c}_{max} < 0.05$ in the volume fraction for all the radial points and load cases but one. The maximum of difference between experimental and calculated value of $\bar{c}$ is $\overline{\Delta c}_{max}$ =0.132 at the edge of a sample for the case $\bar{p}_{max}$= 5.58 GPa.

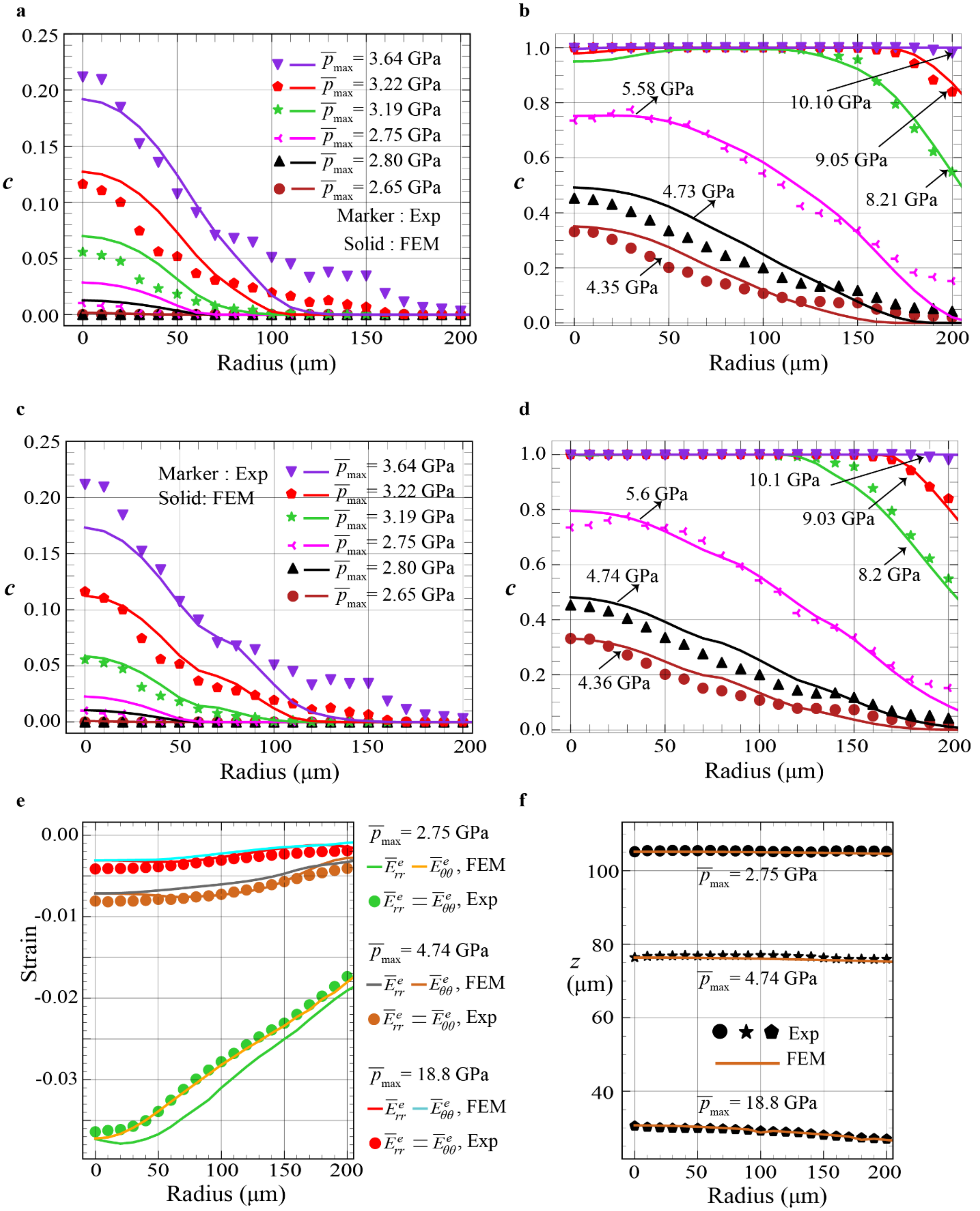

**Fig. 2 Comparison of simulated and experimental radial distributions of various parameters.** **a** and **b** Comparison of $\bar{c}$ for the loadings with lower and higher maximum pressure $\bar{p}_{max}$ at the symmetry axis averaged over the thickness, respectively, for the model with $\delta_1 = 12$, $\delta_2 = 0.925$, $\delta_3 = \delta_5 = 0$, $k = 0.75$, and $\delta_4 = p_\varepsilon^d = 2.65$ GPa and minimized error *Er* = 1.19. **c** and **d** Comparison of $\bar{c}$ for the loadings with lower and higher $\bar{p}_{max}$, respectively, for the model with $\delta_1 = \delta_3 = 0, \delta_2 = 0.803$, $k = 5.20$, and $p_\varepsilon^d = 2.65 - (q_0 - 0.42)$ and minimized error $\overline{\Delta c}_{max}$= 0.071. **e** Comparison of elastic radial $\bar{E}_{rr}^e(r)$ and the hoop $\bar{E}_{\theta\theta}^e(r)$ strains in a mixture averaged over the sample thickness. **f** The sample thickness profiles from the X-ray absorption and FEM.

(b) Much smaller $\overline{\Delta c}_{max}$ with smaller *Er* can be reached within a simpler model with $\delta_1 = \delta_3 = 0$ if we weaken the strict statement in[8,10] that $p_\varepsilon^d$ is getting steady and independent of plastic strain, and assume weak linear dependence of $p_\varepsilon^d$ on $q_0$ with small non-zero $\delta_5$. Minimization of the *Er* for such model resulted in $k = 5.0$, $\delta_2 = 0.775$, $\delta_1 = \delta_3 =$ 0.0, $\delta_4 = 3.07$, $\delta_5 =$ -1.0 with *Er* =0.869 and $\overline{\Delta c}_{max}$ =0.08. If instead of *Er* we minimize $\overline{\Delta c}_{max}$, we obtain $k = 5.2$, $\delta_2 = 0.803$, $\delta_1 = \delta_3 =$ 0.0, $\delta_4 = 3.07$, $\delta_5 =$ -1.0 with close *Er* = 0.871 but slightly better $\overline{\Delta c}_{max}$ =0.07. Figs. 2c,d and Supplementary Fig. S4 show comparison of the calculated and experimental radial distributions of $\bar{c}$. Fig. 3 presents a comparison between experiments and FEA for $\bar{c}$ versus $\bar{p}$ for different radii for these two cases.

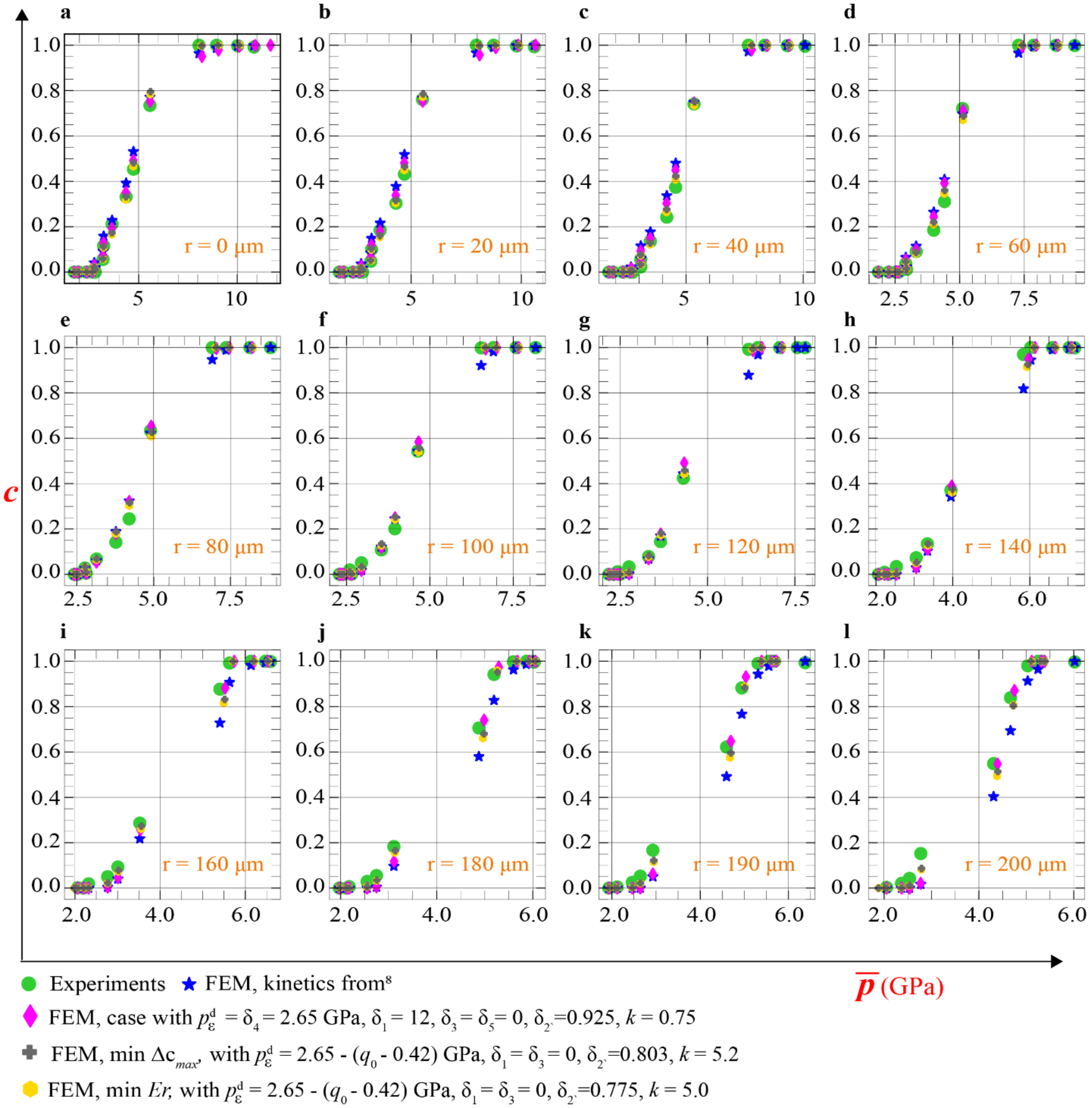

**Fig. 3 Comparison between FEM and experimental volume fractions $\bar{c}$ averaged over the sample thickness versus $\bar{p}$ for different radii.** Small horizontal shifts of points with respect to experiments show small pressure differences for different models.

The fact that we claimed in[8,10] that $p_\varepsilon^d$ is independent of $\boldsymbol{E}^p$ and $\boldsymbol{E}^p$ path, and here we used a weak linear dependence of $p_\varepsilon^d$ on $q_0$ is not contradictory. Independence of $p_\varepsilon^d$ of $\boldsymbol{E}^p$ and $\boldsymbol{E}^p$ path in[8,10] is a correct statement to within some scatter, based on data averaged over the sample thickness, and also should be understood asymptotically with increasing plastic strain. Much more precise method with a weak linear dependence of $p_\varepsilon^d$ on $q_0$ allowed a better correspondence with much larger data set. Still, for some applications this dependence of $p_\varepsilon^d$ on $q_0$ can be neglected. Also, minimum pressure for strain-induced PTs is determined by the dislocation pileup with the largest stress concentrator (i.e., with largest number of dislocations *N*), i.e., by tail in distribution of N in different pileups in the representative volume. Maximum local *N* may reach steady state at larger plastic strain than the average value.

Figs. 2e,f demonstrate a good correspondence between experiments and FEM simulations for the thickness profile of the sample, and radial and azimuthal elastic strains averaged over the sample thickness. This shows that not only PT kinetics but also stresses and elastic strains and vertical displacements are well described, and they are mutually consistent. Results for 3 above sets of material parameters are very close and are shown for the model with minimum $\overline{\Delta c}_{max}$ =0.07 only.

Due to plastic incompressibility conditions, three of four components of the plastic strain tensor, normal $E_{rr}^{p}$, $E_{zz}^{p}$, and shear $E_{rz}^{p}$ are independent. Figs. 1c,d (from the start of loading to the end of PT) and Supplementary Fig. S5 (from the end of PT to the end of the loading) show two 2D projections of the 3D straining trajectory for selected material points. Variation of pressure along each path is also given. One can see a broad variety of magnitudes and loading paths in 4D space $E_{rr}^{p}$, $E_{zz}^{p}$, $E_{rz}^{p}$, and $p$ before, during, and after the PT. While straining trajectories after PT are irrelevant for the current study, they justify the large varieties of $\boldsymbol{E}^{p}$ and $\boldsymbol{E}^{p}$ paths for which rules obtained in[11] (i.e., pressure-dependent yield strength, dislocation density, and crystallite size are independent of $\boldsymbol{E}^{p}$and $\boldsymbol{E}^{p}$ path) are valid. Points along the symmetry axis and plane have zero shear strain, as expected. Trajectories with maximum plastic shear strain are located at the contact surface. Large shear strain generates tensile strain $E_{p,zz}$ because the length of the vertical lateral sides of a material cube increases after its shearing to the parallelopiped with the same height.

Thus, the robust kinetic equation for α-ω PT in Zr is determined, which depends on accumulated plastic strain and pressure and is independent of plastic strain and deviatoric stress tensors. It is valid for a broad range of 3D plastic strain tensor and pressure magnitudes and loading paths, as well as 2D pressure and accumulated plastic strain loading paths (shown in Supplementary Fig. S6a for selected material points). This means that it can be applied for modeling and analyses of various processes involving strain-induced PTs under high pressure.

Evolution of the fields of components of the tensor $\boldsymbol{E}^{p}$, $c$, $q$, and $p$ in the sample is presented in Fig. 4; evolution of the fields of components of the $\boldsymbol{\sigma}$ is shown in Supplementary Fig. S3; the 2D volume fraction evolution versus accumulated plastic strain superposed by pressure for various selected material points is shown in Supplementary Fig. S6b; none of them can be measured experimentally. Unexpectedly, local maximum of plastic strain $E_{rr}^{p}$ is observed at the center of sample, which results in the maximum of $q$. In combination with the maximum normal stresses and therefore, pressure at the center of sample, PT starts there. While there are other zones with larger local maximum of $q$ at the contact surface close to the edge of the diamond culet and at the symmetry plane at the end of the culet, pressures in them are initially well below the minimum PT pressure $p_{\varepsilon}^{d}$, and therefore PT does not start. With further loading, these regions partially or completely flow outside the culet region not studied experimentally. With further compression, ω-Zr zone grows from the center to the periphery and contact surfaces anisotropically, and volume fraction in the zone increases. Much faster growth towards the periphery is partially due to radial flow of the already transformed regions. Evidently, very heterogeneous distributions of $q$ and $c$ along the thickness at the symmetry axis is obtained, violating the main assumption for the derivation of the kinetic equation in[8,10].

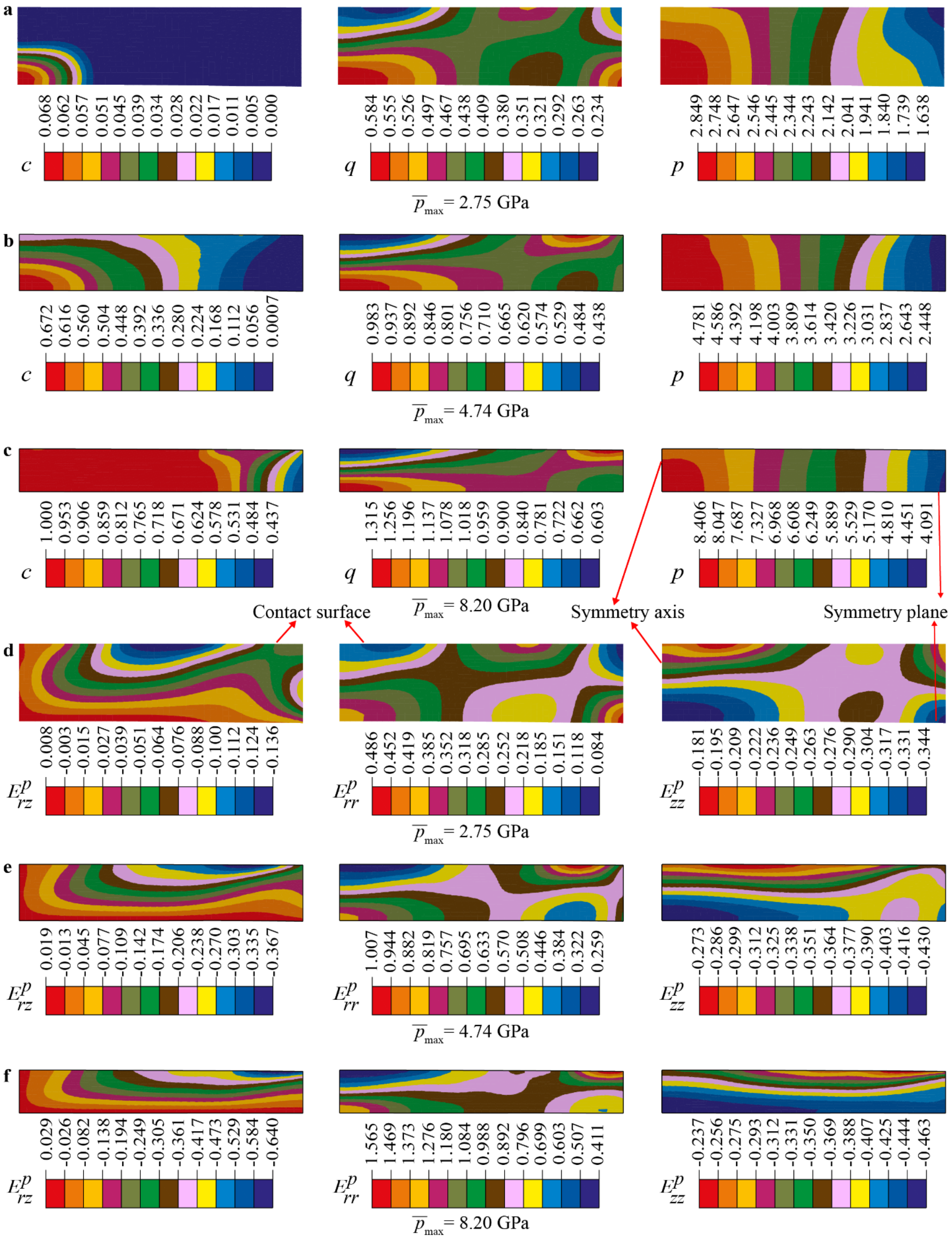
a
0.068 0.062 0.057 0.051 0.045 0.039 0.034 0.028 0.022 0.017 0.011 0.005 0.000
c
0.584 0.555 0.526 0.497 0.467 0.438 0.409 0.380 0.351 0.321 0.292 0.263 0.234
q
2.849 2.748 2.647 2.546 2.445 2.344 2.243 2.142 2.041 1.941 1.840 1.739 1.638
p
$\bar{p}_{max}$ = 2.75 GPa
b
0.672 0.616 0.560 0.504 0.448 0.392 0.336 0.280 0.224 0.168 0.112 0.056 0.0007
c
0.983 0.937 0.892 0.846 0.801 0.756 0.710 0.665 0.620 0.574 0.529 0.484 0.438
q
4.781 4.586 4.392 4.198 4.003 3.809 3.614 3.420 3.226 3.031 2.837 2.643 2.448
p
$\bar{p}_{max}$ = 4.74 GPa
c
1.000 0.953 0.906 0.859 0.812 0.765 0.718 0.671 0.624 0.578 0.531 0.484 0.437
c
1.315 1.256 1.196 1.137 1.078 1.018 0.959 0.900 0.840 0.781 0.722 0.662 0.603
q
8.406 8.047 7.687 7.327 6.968 6.608 6.249 5.889 5.529 5.170 4.810 4.451 4.091
p
$\bar{p}_{max}$ = 8.20 GPa
Contact surface
Symmetry axis
Symmetry plane
d
0.008 -0.003 -0.015 -0.027 -0.039 -0.051 -0.064 -0.076 -0.088 -0.100 -0.112 -0.124 -0.136
$E^p_{rz}$
0.486 0.452 0.419 0.385 0.352 0.318 0.285 0.252 0.218 0.185 0.151 0.118 0.084
$E^p_{rr}$
-0.181 -0.195 -0.209 -0.222 -0.236 -0.249 -0.263 -0.276 -0.290 -0.304 -0.317 -0.331 -0.344
$E^p_{zz}$
$\bar{p}_{max}$ = 2.75 GPa
e
0.019 -0.013 -0.045 -0.077 -0.109 -0.142 -0.174 -0.206 -0.238 -0.270 -0.303 -0.335 -0.367
$E^p_{rz}$
1.007 0.944 0.882 0.819 0.757 0.695 0.633 0.570 0.508 0.446 0.384 0.322 0.259
$E^p_{rr}$
-0.273 -0.286 -0.299 -0.312 -0.325 -0.338 -0.351 -0.364 -0.377 -0.390 -0.403 -0.416 -0.430
$E^p_{zz}$
$\bar{p}_{max}$ = 4.74 GPa
f
0.029 -0.026 -0.082 -0.138 -0.194 -0.249 -0.305 -0.361 -0.417 -0.473 -0.529 -0.584 -0.640
$E^p_{rz}$
1.565 1.469 1.373 1.276 1.180 1.084 0.988 0.892 0.796 0.699 0.603 0.507 0.411
$E^p_{rr}$
-0.237 -0.256 -0.275 -0.293 -0.312 -0.331 -0.350 -0.369 -0.388 -0.407 -0.425 -0.444 -0.463
$E^p_{zz}$
$\bar{p}_{max}$ = 8.20 GPa

**Fig. 4 Distributions of volume fraction of ω-Zr $c$, accumulated plastic strain $q$, pressure $p$ and components of Lagrangian plastic strains in a sample for three loadings characterized by $\bar{p}_{max}$ for the case with parameters $\delta_1 = \delta_3 = 0, \delta_2 = 0.803, k = 5.20$, and $p_\varepsilon^d = 2.65 - (q_0 - 0.42)$ in the kinetic equation.**

Note that since we found that for each radius PT starts at the symmetry plane, where shears are zero, independence of $p_\varepsilon^d$ of $\boldsymbol{E}^p$ and straining path can be strictly claimed for zero shears only. However, due to averaging of experimental data over the thickness, and since away from the center, the boundary between transformed and non-transformed regions are close to vertical, data from the regions with shears contribute as well. This is not the case for the symmetry axis. Our results suggest that the statement in[8,10] of the independence of $p_\varepsilon^d$ of the plastic strain tensor and its path after severe plastic deformations requires further justification with larger strains and shear components. Torsion in rotational DAC can significantly extend classes of straining paths. Also, one may need to distinguish between the relatively small increment of monotonous deformation in DAC and preliminary severe deformation by rolling at normal pressure. While rolling has already produced steady crystallite size, dislocation density, and hardness, under high-pressure compression in DAC, another steady state was reached with smaller steady crystallite size and larger dislocation density[11].

To summarize, a CE-FEM method is developed including (a) FEM modeling of α-ω PT and coupled plastic flow; (b) experimental input in terms of pressure-dependent elastoplastic properties and friction condition between sample and diamond, and (c) determination of $\boldsymbol{E}^p$, $\boldsymbol{\sigma}$, and $c$ fields and kinetic equation for strain-induced α-ω PT in a strongly plastically predeformed Zr sample. This is done by iteratively solving an inverse problem on determination of 5 material parameters in the physics-based kinetic equation. Very good correspondence between calculated and experimental curves is observed for the volume fraction of ω-Zr and radial and azimuthal elastic strains averaged over the sample thickness, as well as thickness profile of the sample. Obtained broad variety of magnitudes and loading paths in 4D space $E_{rr}^p$, $E_{zz}^p$, $E_{rz}^p$, and $p$ in a sample proves that the kinetic equation is independent of plastic strain and deviatoric stress tensors and also justifies the large varieties of $\boldsymbol{E}^p$ and $\boldsymbol{E}^p$ paths for which rule obtained in[11] (i.e., that pressure-dependent yield strength, dislocation density, and crystallite size are independent of $\boldsymbol{E}^p$ and $\boldsymbol{E}^p$ path) are valid. We found that $p_\varepsilon^d$ is a weakly decreasing linear function of $q$; this refines the rule found in[8,10] that with increasing plastic strain and within some scatter, $p_\varepsilon^d$ is asymptotically independent of $q$, $\boldsymbol{E}^p$, and $\boldsymbol{E}^p$ path. Since it is found in[30] that during α-ω PT in Zr, dislocation density and crystalline size are unique functions of the volume fraction of ω-Zr independent of $\boldsymbol{E}^p$ and $\boldsymbol{E}^p$ path, and $p$, we can immediately obtain explicit kinetics for these parameters using Eq. (1). Similar refinement can be produced for the pressure-dependent yield strength, dislocation density, and crystallite size of ω phase in[11]. Obtained results initiate quantitative kinetic studies of strain-induced PTs and promise to bring efforts in the above fields to a qualitatively higher level. Similar methods can be applied for other materials, and also extended for annealed materials and high strain rates, and for finding kinetics for other parameters, like dislocation density and crystallite size, including materials without PTs. Kinetic equation of the type of Eq. (1) can be used for quantitative modeling and optimization of the processes involved in (a) defect-induced synthesis of nanostructured materials, phases, and nanocomposites by severe plastic deformation with high-pressure torsion[28] and ball milling[2], (b) friction and wear[3,12,13], (c) surface processing (polishing, turning, scratching, etc.)[29], (d) high-pressure geology (mechanism of the deep-focus earthquakes, microdiamond appearance, and study of multiple PTs during plastic flow, which are currently described as pressure-induced)[1,3,15,19], and (e) astrogeology[20,21].

## METHODS

### FEM modeling and simulations

A large elastoplastic strain model for mixture of α- and ω-Zr using the mixture rule for all properties coupled with modeling plastic-strain induced PT kinetics is advanced (see Supplementary Notes). The characterization of the kinetic equation (1) is done by iteratively running the FEM simulations such that the difference between experimental and simulated volume fraction minimizes the residual errors at all loads and radial points. The robustness of the FEM model is enhanced by using the shear stress at the sample-diamond contact surface from the CEA method in ref. (*8*). In the culet portion, the shear stress from CEA is used as the boundary condition for modeling contact between sample and the diamond. At the inclined portion of the sample-anvil contact surface, the critical friction stress is determined by the minimum between the shear stress from the CEA method and Coulomb friction. The elastic constitutive response of polycrystalline Zr is modeled using 3$^{rd}$ order Murnaghan potential. Pressure-dependent *von-Mises* yield equation for isotropic perfectly plastic polycrystal is used to model the plastic response of the sample. Associated flow rule in deviatoric stress space is used along with plastic incompressibility. Kinematic compatibility is enforced by decomposing the total deformation gradient multiplicatively into elastic, plastic, and transformation components. The elastic response of the diamond is modeled using 4$^{th}$ order elastic potential for cubic crystal averaged over azimuthal direction to keep the axial symmetry.

### Materials

The material studied in this paper is the same as that used by Zhilyaev et al.[31] and was purchased from Haines and Maassen (Bonn, Germany). It is commercially pure $\alpha$-Zr (with Fe: 330 ppm; Mn: 27 ppm; Hf: 452 ppm; S: <550 ppm; Nd: < 500 ppm). The initial sample slab, with a thickness of 5.25 mm, was cold rolled to achieve a final thickness of approximately 165 μm, resulting in a plastically pre-deformed sample with saturated hardness. The Vickers microhardness test method was utilized to characterize the hardness of the sample at various stages during cold rolling. A 3 mm diameter disk was punch-cut from the obtained thin-rolled sheet for unconstrained compression experiments in a diamond anvil cell (DAC). For hydrostatic compression experiments, small specks approximately 20 μm in size were chipped off from the plastically pre-deformed sample using a diamond file.

The hydrostatic high-pressure X-ray diffraction measurements were conducted to estimate the equation of state, bulk modulus, and its pressure derivative at ambient pressure using the same DAC as was used for non-hydrostatic DAC experiments. For these experiments, small Zr specks of approximately 20 μm in size, as already mentioned, were loaded into the sample chamber along with silicone oil and copper chips, serving as the pressure transmitting medium and pressure marker, respectively. The sample chamber was prepared by drilling a hole ~ 250 μm in diameter in steel gaskets pre-indented using diamond anvils, reducing their initial thickness from ~ 250 μm to about 50 μm. Hydrostatic high-pressure experiments were conducted in small pressure increments of around 0.2 GPa, up to a maximum pressure of 16 GPa.

**Experimental techniques and methodology**

Unconstrained plastic compression experiments were conducted by applying various compression loads to a plastically pre-deformed Zr sample loaded into the DAC without the use of any constraining gasket. The sample was subjected to an axial load ranging from 50 N to the maximum of 1000 N with 26 increments.

In situ XRD experiments were performed at the 16-BM-D beamline of HPCAT, Sector 16 at the Advanced Photon Source, utilizing focused monochromatic X-rays with a wavelength of 0.3096(3) Å and a size of approximately 6 μm × 5 μm (full width at half maximum (FWHM)). For each load-condition, the sample was radially scanned across the entire culet diameter (500 μm) in steps of 10 μm, and 2D diffraction images were recorded using a Perkin Elmer flat panel detector. At each load step, an X-ray absorption scan was also recorded in the same 10 μm steps to obtain the thickness profile of the sample under the given load condition, see details in[10].

The 2D diffraction images were converted to a 1D diffraction pattern using FIT2D software[32,33], and subsequently analyzed through Rietveld refinement[34,35] using GSAS II[36] and MAUD[37] software. This analysis aimed to obtain lattice parameters, phase fractions and texture parameters for both α and ω- Zr. The different angular dependence of the grain size and microstrain contributions to the diffraction peak broadening allows for their separation. We employed a whole powder pattern fitting approach using the modified Rietveld method, as implemented in MAUD software[38], which accounts for texturing and stress anisotropy.

In axial geometry (i.e., when the incident X-ray beam is directed along z-axis) (Fig. 1a), diffraction condition is satisfied primarily for crystallographic planes that are nearly parallel (with plane normal perpendicular) to the load axis. Therefore, the observed shifts in diffraction peaks can be practically utilized to estimate strains in the radial and azimuthal directions viz. $\bar{E}_{11}^{e} = \bar{E}_{rr}^{e}$ and $\bar{E}_{22}^{e} = \bar{E}_{\theta\theta}^{e}$ averaged over the sample thickness. Ideally, the angle between the load axis and diffraction vector ($\psi$) should be 90° to accurately estimate these strain components. However, since achieving a 90° angle between the load axis and the diffraction vector is not possible in axial geometry, we can instead use the diffraction peak with the smallest diffraction angle, $\theta$. In our experiments with $\alpha$-Zr, the (100) diffraction peak appears at $\theta = 3.18°$ for the X-rays used ($\lambda$ = 3.1088 Å) at ambient pressure. This corresponds to $\psi = 86.82°$ and can be used to estimate the strain components $\bar{E}_{rr}^{e}$ and $\bar{E}_{\theta\theta}^{e}$. Note that the (100) peak corresponds to the $a$ lattice parameter because the $c$-axis of $\alpha$-Zr is predominantly aligned along the loading direction, as determined by our texture analysis.

For ω-Zr, the (001) diffraction peak appears at $\theta = 2.85°$ ($\psi = 87.15°$) and can be used to estimate the strain components $\bar{E}_{rr}^{e}$ and $\bar{E}_{\theta\theta}^{e}$. The (001) peak of ω-Zr corresponds to the $c$ lattice parameter, and according to texture analysis, the $c$-axis of ω-Zr is predominantly perpendicular to the loading direction of the DAC.

Thus, strain components $\bar{E}^e_{rr}$ and $\bar{E}^e_{\theta\theta}$ for α and ω phases of Zr have been obtained for each loading condition at each scanning position using the following equations:

For $\alpha$-Zr:

$\bar{E}^e_{rr} = 0.5\left(\left(a_{\phi=0^o}/a_0\right)^2 - 1\right)$ using $\phi = 0^o$ sector of (100) diffraction ring;

$\bar{E}^e_{\theta\theta} = 0.5\left(\left(a_{\phi=90^o}/a_0\right)^2 - 1\right)$ using $\phi = 90^o$ sector of (100) diffraction ring;

For $\omega$-Zr :

$\bar{E}^e_{rr} = 0.5\left(\left(c_{\phi=0^o}/c_0\right)^2 - 1\right)$ using $\phi = 0^o$ sector of (001) diffraction ring;

$\bar{E}^e_{\theta\theta} = 0.5\left(\left(c_{\phi=90^o}/c_0\right)^2 - 1\right)$ using $\phi = 90^o$ sector of (001) diffraction ring.

## DATA AVAILABILITY

Source data are provided with this paper.

## REFERENCES

1. Green, H. W. & Burnley, P. C. A new, self-organizing, mechanism for deep-focus earthquakes. *Nature* **341**, 773–737 (1989).
2. Maria Carta, Leonarda Vugrin, Goran Miletić, Marina Juribašić Kulcsár, Pier Carlo Ricci, Ivan Halasz,* and Francesco Delogu. Mechanochemical Reactions: from Individual Impacts to Global Transformation Kinetics. Angewandte Chemie Int. Ed. 2023, 62, e202308046.
3. Green II, H. W., Shi, F., Bozhilov, K., Xia, G. & Reches, Z. Phase transformation and nanometric flow cause extreme weakening during fault slip. Nat. Geosci. 8, 484–489 (2015).
4. Chen, H., Levitas, V. I. & Xiong, L. Amorphization induced by 60$^o$ shuffle dislocation pileup against tilt grain boundaries in silicon bicrystal under shear. *Acta Mater.* **179**, 287–295 (2019).
5. Javanbakht, M. & Levitas, V. I. Phase field simulations of plastic strain-induced phase transformations under high pressure and large shear. *Phys. Rev. B* **94**, 214104 (2016).
6. Zharov, A. *Reaction of Solid Monomers and Polymers under Shear Deformation and High Pressure. High Pressure Chemistry and Physics of Polymers chapter 7, ed. by Kovarskii*. (Boca Raton, CRC Press, 1994 pp. 267301).
7. Edalati, K., Horita, Z., Yagi, S. & Matsubara, E. Allotropic phase transformation of pure zirconium by high-pressure torsion. *Mat. Sci. Eng. A* **523**, 277–281 (2009).
8. Levitas, V. I., Dhar, A. & Pandey, K. K. Tensorial stress-plastic strain fields in $\alpha$- $\omega$ Zr mixture, transformation kinetics, and friction in diamond anvil cell. *Nature Communication* **14**, 5955 (2023).

9. Levitas, V. I., Esfahani, S. E. & Ghamarian, I. Scale-free modeling of coupled evolution of discrete dislocation bands and multivariant martensitic microstructure. *Phys. Rev. Lett.* **121**, 205701 (2018).
10. Pandey, K. K. & Levitas, V. I. In situ quantitative study of plastic strain-induced phase transformations under high pressure: Example for ultra-pure Zr. *Acta Materialia* **196**, 338-346 (2020).
11. Lin, F., Levitas, V. I., Pandey, K. K., Yesudhas, S. & Park, C. In-situ study of rules of nanostructure evolution, severe plastic deformations, and friction under high pressure. *Materials Research Letters* **11**, 757-763 (2023).
12. Pastewka, L., Moser, S., Gumbsch, P. & Moseler, M. Anisotropic mechanical amorphization drives wear in diamond. *Nature Materials* **10(1)**, 34–38 (2011).
13. Reichenbach, T., Moras, G., Pastewka, L. & Moseler, M. Solid-phase silicon homoepitaxy via shear-induced amorphization and recrystallization. *Physical Review Letters* **127(12)**, 126101 (2021).
14. Ji, C., Levitas, V. I., Zhu, H., Chaudhuri, J., Marathe, A. & Ma, Y. Shear-induced phase transition of nanocrystalline hexagonal Boron Nitride to Wurtzitic structure at room temperature and low pressure. *Proceedings of the National Academy of Sciences of the United States of America* **109**, 19108-19112 (2012).
15. Gao, Y., Ma, Y., An, Q., Levitas, V., Zhang, Y., Feng, B., Chaudhuri, J. & Goddard III, W. A. Shear driven formation of nano-diamonds at sub-gigapascals and 300 K. *Carbon* **146**, 364-368 (2019).
16. Levitas, V. I. High-pressure mechanochemistry: Conceptual multiscale theory and interpretation of experiments. *Phys. Rev. B* **70**, 184118 (2004).
17. Levitas, V. I., Ma, Y., Selvi, E., Wu, J. & Patten, J. High-density amorphous phase of silicon carbide obtained under large plastic shear and high pressure, *Phys. Rev. B* **85**, 054114 (2012).
18. Blank, V. D. & Estrin, E. I. *Phase Transitions in Solids under High Pressure* (New York, CRC Press, 2014).
19. Levitas, V. I. Resolving puzzles of the phase-transformation-based mechanism of the deep-focus earthquake. *Nature Communications* **13**, 6291 (2022).
20. Edalati, K., Taniguchi, I., Floriano, R. & Luchessi, A. D. Glycine amino acid transformation under impacts by small solar system bodies, simulated by high-pressure torsion method, *Sci. Rep.* **12** 5677 (2022).
21. Stolar, T., Grubešić, S., Cindro, N., Meštrović, E., Užarević, K. & Hernández, J. G. Mechanochemical Prebiotic Peptide Bond Formation. *Angewandte Chemie International Edition* **60**, 12727-12731 (2021).
22. Levitas, V.I. & Zarechnyy, O.M. Modeling and simulation of strain-induced phase transformations under compression in a diamond anvil cell. *Physical Review B* **82**, 174123 (2010).
23. Levitas, V.I. & Zarechnyy, O.M. Modeling and simulation of strain-induced phase transformations under compression and torsion in a rotational diamond anvil cell. *Physical Review B* **82**, 174124 (2010).

24. Feng, B. & Levitas, V. I. Coupled elastoplasticity and plastic strain-induced phase transformation under high pressure and large strains: Formulation and application to BN sample compressed in a diamond anvil cell. *Int. J. Plast.* **96**, 156-181 (2017).
25. Feng, B., Levitas, V. I. & Li, W. FEM modeling of plastic flow and strain-induced phase transformation in BN under high pressure and large shear in a rotational diamond anvil cell. *Int. J. Plast.* **113**, 236–254 (2019).
26. Levitas, V. I. High-pressure phase transformations under severe plastic deformation by torsion in rotational anvils. *Material Transactions* **60**, 1294-1301 (2019).
27. Levitas, V. I. Recent in situ Experimental and Theoretical Advances in Severe Plastic Deformations, Strain-Induced Phase Transformations, and Microstructure Evolution under High Pressure. *Material Transactions*. https://doi.org/10.2320/matertrans.MT-MF2022055 (2023).
28. Edalati, K., Bachmaier, A., Beloshenko, V. A., et al. Nanomaterials by Severe Plastic Deformation: Review of Historical Developments and Recent Advances. *Materials Research Letters* **10**, 163-256 (2022).
29. Patten, J. A., Cherukuri, H. & Yan, J. Ductile-regime machining of semiconductors and ceramics, *In High-Pressure Surface Science and Engineering* 543-632 (CRC Press, 2019).
30. Lin, F., Levitas, V. I., Pandey, K. K., Yesudhas, S. & Park, C. Rules of plastic strain-induced phase transformations and nanostructure evolution under high-pressure and severe plastic flow. Preprint at https://doi.org/10.48550/arXiv.2305.15737 (2023).
31. Zhilyaev, A., Gálvez, F., Sharafutdinov, A. & Pérez-Prado, M. Influence of the high-pressure torsion die geometry on the allotropic phase transformations in pure Zr. *Mat. Sci. Eng. A* **527**, 3918-3928 (2010).
32. Hammersley, A. Fit2d: An introduction and overview, in ESRF Internal Report, ESRF97HA02T (Institute of Physics, Bristol, 1997).
33. Hammersley, A., Svensson, S., Hanfland, M., Fitch, A. & Hausermann, D. Two-dimensional detector software: From real detector to idealised image or two-theta scan. *High Press. Res.* **14**, 235–248 (1996).
34. Rietveld, H. A profile refinement method for nuclear and magnetic structures. *J. Appl. Cryst.* **2**, 65–71 (1969).
35. Young, R. The Rietveld Method, (International Union of Crystallography, Oxford University Press, 1993).
36. Toby, B. & Von Dreele, R. GSAS-II: the genesis of a modern opensource all-purpose crystallography software package, *J. Appl. Cryst.* **46**, 544–549 (2013).
37. Ferrari, M. & Lutterotti, L. Method for the simultaneous determination of anisotropic residual stresses and texture by X-ray diffraction. *J. Appl. Phys.* **76**, 7246-55 (1994).
38. Lutterotti, L. Total pattern fitting for the combined size–strain–stress–texture determination in thin film diffraction. *Nuclear Inst. and Methods in Physics Research B,* **268**, 334-340 (2010).
39. Supplementary Information

## ACKNOWLEDGMENTS

Support from NSF (CMMI-1943710, DMR-2246991, and XSEDE MSS170015) and Iowa State University (Vance Coffman Faculty Chair Professorship and Murray Harpole Chair in Engineering) for VIL and AD is greatly appreciated. AD also acknowledges support from INTERN supplement to NSF grant CMMI-1943710 and from HPCAT for an internship at HPCAT. KKP was supported by NSF (CMMI-1943710) and ISU. NV work performed under the auspices of the U.S. Department of Energy by Lawrence Livermore National Laboratory under Contract DE-AC52-07NA27344. Portions of this work were performed at HPCAT (Sector 16), Advanced Photon Source (APS), Argonne National Laboratory. HPCAT operations are supported by DOE-NNSA's Office of Experimental Sciences. The Advanced Photon Source is a U.S. Department of Energy (DOE) Office of Science User Facility operated for the DOE Office of Science by Argonne National Laboratory under Contract No. DE-AC02-06CH11357.

## AUTHOR CONTRIBUTIONS

VIL conceived the study, supervised the project, developed theoretical models, and secured funding. AD performed the simulations. AD and VIL prepared initial manuscript. KKP performed experiments, collected and postprocessed data. CP set up the synchrotron XRD experiments. All authors contributed to discussions of the data and to the writing of the manuscript.

## COMPETING INTERESTS

The authors declare no competing interests.

## SUPPLEMENTARY INFORMATION

Supplementary Information is available for this paper.

Supplementary Information

**Quantitative kinetic rules for plastic strain-induced $\alpha$ - $\omega$ phase transformation in Zr under high pressure**

Achyut Dhar[1,5*], Valery I. Levitas[1,2*], K.K. Pandey[1,3,4], Changyong Park[5], Maddury Somayazulu[5], Nenad Velisavljevic[5,6]

[1] Department of Aerospace Engineering, Iowa State University, Ames, IA 50011, USA
[2] Department of Mechanical Engineering, Iowa State University, Ames, IA 50011, USA
[3] High Pressure and Synchrotron Radiation Physics Division, Bhabha Atomic Research Centre, Bombay, Mumbai-400085, India (present affiliation)
[4] Homi Bhabha National Institute, Anushaktinagar, Mumbai 400094, India (present affiliation)
[5] HPCAT, X-ray Science Division, Argonne National Laboratory, Argonne, Illinois 60439, USA
[6] Lawrence Livermore National Laboratory, Physics Division, Livermore, CA 94550, USA

* Corresponding authors: Email: vlevitas@iastate.edu and adhar@iastate.edu
These authors contributed equally: Valery I. Levitas, Achyut Dhar

**This PDF file includes:**

# Supplementary Notes

## 1 A combined experimental-computational approach for finding fields of all components of stress and plastic strain tensors, and volume fraction of phases in a Zr sample compressed in DAC

The flowchart of the interaction between different methods is presented in Supplementary Fig. 1.

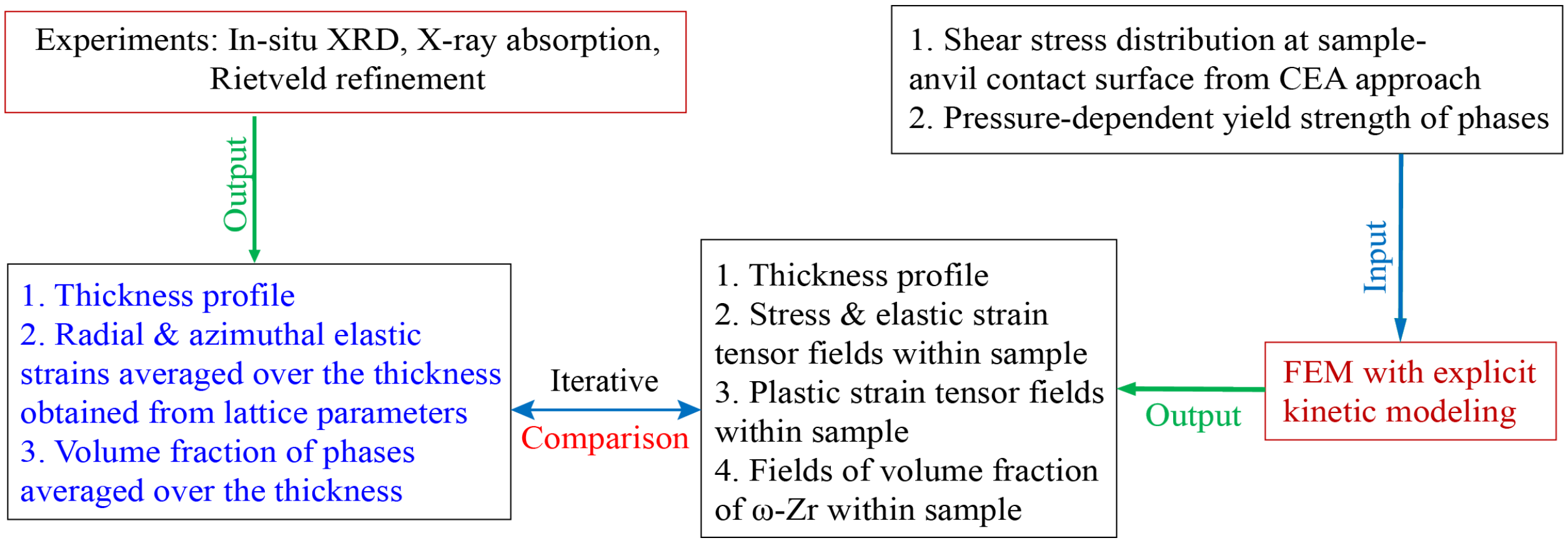

Supplementary Fig. S1: **The flowchart of the interaction between experimental and FE methods.**

Experimental methods based on in-situ X-ray diffraction and absorption allow us to determine the radial distributions of the elastic radial $\bar{E}_{0,rr}(r)$ and azimuthal $\bar{E}_{0,\theta\theta}(r)$ strains in $\alpha$ and $\omega$-Zr phases, volume fraction of $\omega$ phase $\bar{c}(r)$ (bar over the field variables means averaged over the sample thickness), as well as sample thickness profile $h(r)$. The friction stress distribution from the analytical model [1] is utilized as the boundary condition in our FEM problem formulation along with the elastoplastic and transformation properties, as well as the pressure-dependent yield strengths of $\alpha-$ & $\omega$-Zr. FEM solution delivers all components of stress, elastic, and plastic strain tensors, the sample thickness profile, and the field of volume fraction in the entire sample. FEM-based elastic radial $\bar{E}_{0,rr}$ and azimuthal $\bar{E}_{0,\theta\theta}$ strain distributions, volume fraction $\bar{c}$ distributions, and the sample thickness profile are compared with experiments to validate FEM modeling, and, consequently, the entire procedure and all fields.

## 2 FEM simulations

### 2.1 Complete system of equations for FEM simulations [2]

Box 1 summarizes all mechanical equations derived in [2] in the form used in our simulations and also the physics-based plastic-strain induced kinetic equation. Vectors and tensors are represented in boldface type, e.g., $\boldsymbol{A} = A_{ij}\boldsymbol{e}_i\boldsymbol{e}_j$, where $A_{ij}$ are the components in the Cartesian system with unit basis vectors $\boldsymbol{e}_i$ and summation over repeated indices is assumed. The expression $\boldsymbol{e}_i\boldsymbol{e}_j$ denotes the direct or dyadic product of vectors, representing the second-rank tensor. Let $\boldsymbol{A}\cdot\boldsymbol{B} = A_{ik}B_{kj}\boldsymbol{e}_i\boldsymbol{e}_j$ and $\boldsymbol{A}:\boldsymbol{B} = tr(\boldsymbol{A}\cdot\boldsymbol{B}) = A_{ij}B_{ji}$ denote the contraction (or scalar product) of tensors over first and second nearest indices, respectively. Here, $tr$ represents the trace operation (the sum of the diagonal components), and $A_{ik}B_{kj}$ denotes the matrix product. Additionally, $\boldsymbol{A}_s =: \frac{\boldsymbol{A}+\boldsymbol{A}^t}{2}$ and $\boldsymbol{A}_a =: \frac{\boldsymbol{A}-\boldsymbol{A}^t}{2}$ are respectively the symmetric and anti-symmetric components of $\boldsymbol{A}$, where $'t'$ in the superscript designates the transpose operation defined as $\boldsymbol{A}^t = A_{ji}\boldsymbol{e}_i\boldsymbol{e}_j$, given $\boldsymbol{A} = A_{ij}\boldsymbol{e}_i\boldsymbol{e}_j$.

**Box 1. The complete system of equations**

*The multiplicative decomposition of the total deformation gradient* $\boldsymbol{F}$ *into elastic* $\bar{\boldsymbol{F}}_e$ *and inelastic* $\boldsymbol{F}_i$ *contributions:*

$$\begin{aligned}\boldsymbol{F} = \frac{\partial \boldsymbol{r}}{\partial \boldsymbol{r}_0} &= \bar{\boldsymbol{F}}_e \cdot \boldsymbol{F}_i \\ &= \bar{\boldsymbol{R}}_e \cdot \bar{\boldsymbol{U}}_e \cdot \boldsymbol{R}_i \cdot \boldsymbol{U}_i \\ &= \bar{\boldsymbol{R}}_e \cdot \boldsymbol{R}_i \cdot \boldsymbol{R}_i^t \cdot \bar{\boldsymbol{U}}_e \cdot \boldsymbol{R}_i \cdot \boldsymbol{U}_i \\ &= \boldsymbol{R}_e \cdot \boldsymbol{U}_e \cdot \boldsymbol{U}_i \\ &= \boldsymbol{F}_e \cdot \boldsymbol{U}_i \\ &= \boldsymbol{V}_e \cdot \boldsymbol{R}_e \cdot \boldsymbol{U}_i. \end{aligned} \quad \text{(S1)}$$

$$\boldsymbol{R}_e = \bar{\boldsymbol{R}}_e \cdot \boldsymbol{R}_i; \quad \boldsymbol{U}_e = \boldsymbol{R}_i^t \cdot \bar{\boldsymbol{U}}_e \cdot \boldsymbol{R}_i; \quad \boldsymbol{F}_e = \boldsymbol{R}_e \cdot \boldsymbol{U}_e = \boldsymbol{V}_e \cdot \boldsymbol{R}_e$$
$$\boldsymbol{B}_e = 0.5(\boldsymbol{F}_e \cdot \boldsymbol{F}_e^t - \boldsymbol{I}) = 0.5(\boldsymbol{V}_e^2 - \boldsymbol{I}).$$

Here, $\boldsymbol{r}$ and $\boldsymbol{r}_0$ represent the position vectors of a material point in the current and undeformed configurations, respectively; $\bar{\boldsymbol{R}}_e$ and $\bar{\boldsymbol{U}}_e$ are respectively the elastic right stretch and elastic proper orthogonal tensors from the polar decomposition of $\bar{\boldsymbol{F}}_e$; $\boldsymbol{R}_i$ and $\boldsymbol{U}_i$ are respectively the inelastic right stretch and inelastic proper orthogonal tensors from the polar decomposition of $\boldsymbol{F}_i$; $\boldsymbol{V}_e$ and $\boldsymbol{R}_e$ are respectively the elastic left stretch and elastic proper orthogonal tensors from the polar decomposition of $\boldsymbol{F}_e$; $\boldsymbol{B}_e$ denotes the elastic Eulerian strain tensor and $\boldsymbol{I}$ the unit tensor.

Since transformation strain is assumed to be pure dilatational, we obtain for transformational $\boldsymbol{U}_t$ and plastic $\boldsymbol{U}_p$ deformation right stretch tensors

$$\boldsymbol{U}_t = (1+\bar{\varepsilon}_{t0}c)\boldsymbol{I}; \qquad \boldsymbol{U}_p = \boldsymbol{U}_i/(1+\bar{\varepsilon}_{t0}c); \qquad \boldsymbol{U}_i = \boldsymbol{U}_t \cdot \boldsymbol{U}_p. \quad \text{(S2)}$$

Here, $\bar{\varepsilon}_{t0} = -0.0158$ denotes the volumetric transformation strain for $\alpha \to \omega$ Zr and $c$ is the volume fraction of $\omega-$Zr. We utilize in the main text Lagrangian elastic $\boldsymbol{E}^e$ and plastic $\boldsymbol{E}^p$ strains

$$\boldsymbol{E}^e = 0.5(\boldsymbol{F}_e^t \cdot \boldsymbol{F}_e - \boldsymbol{I}) = \boldsymbol{R}_e^t \cdot \boldsymbol{B}^e \cdot \boldsymbol{R}_e; \qquad \boldsymbol{E}^p = 0.5(\boldsymbol{U}_p \cdot \boldsymbol{U}_p - \boldsymbol{I}), \quad \text{(S3)}$$

where using superscripts for $\boldsymbol{E}^e$ and $\boldsymbol{E}^p$ instead of subscripts is convenient for presenting components of these tensors in the main text.

*The decomposition of the deformation rate* $\boldsymbol{d}$ *into elastic, plastic, and transformation com-*

*ponents:*

$$\boldsymbol{d} = \overset{\nabla}{\boldsymbol{B}}_e \cdot \boldsymbol{V}_e^{-2} + 2(\boldsymbol{d} \cdot \boldsymbol{B}_e)_a \cdot \boldsymbol{V}_e^{-2} + \boldsymbol{\gamma} + \bar{\varepsilon}_{t0}\dot{c}\boldsymbol{I}. \tag{S4}$$

Here, $\overset{\nabla}{\boldsymbol{B}}_e = \dot{\boldsymbol{B}}_e - 2(\boldsymbol{w} \cdot \boldsymbol{B}_e)_s$ denotes the Jaumann time derivative of $\boldsymbol{B}_e$; $\boldsymbol{w}$ is the anti-symmetric part of the velocity gradient in the current configuration, and $\boldsymbol{\gamma}$ signifies the plastic part of the deformation rate.

*The third-order Murnaghan potential:*

$$\psi(\boldsymbol{B}_e) = \frac{\lambda + 2G}{2} I_1^2 - 2GI_2 + \left( \frac{l + 2m}{3} I_1^3 - 2mI_1I_2 + nI_3 \right), \tag{S5}$$

where $\lambda$, $G$, $l$, $m$, and $n$ represent the Murnaghan material constants of the mixture, while $I_1$, $I_2$, and $I_3$ are the invariants of $\boldsymbol{B}_e$, defined as:

$$\begin{aligned} &I_1 = tr(\boldsymbol{B}_e) = B_{e11} + B_{e22} + B_{e33}; \\ &I_2 = \frac{1}{2}\left[(\mathrm{tr}(\mathbf{B}_e))^2 - \mathrm{tr}(\mathbf{B}_e^2)\right] = \frac{1}{2} B_{e22}B_{e33} - B_{e23}^2 + B_{e11}B_{e33} - B_{e13}^2 + B_{e22}B_{e11} - B_{e12}^2; \\ &I_3 = \det(B_e). \end{aligned} \tag{S6}$$

A simple mixture rule is used to obtain the Murnaghan constants of the mixture

$$\begin{aligned} &\lambda = (1-c)\lambda_1 + c\lambda_2; \quad G = (1-c)G_1 + cG_2; \quad m = (1-c)m_1 + cm_2; \\ &l = (1-c)l_1 + cl_2; \quad n = (1-c)n_1 + cn_2. \end{aligned} \tag{S7}$$

Here, the subscripts 1 and 2 designate $\alpha$- and $\omega$-Zr.

*Elasticity rule for the Cauchy (true) stress:*

$$\begin{aligned} \boldsymbol{\sigma} &= J_e^{-1}(2\boldsymbol{B}_e + \boldsymbol{I}) \cdot \frac{\partial \psi}{\partial \boldsymbol{B}_e} \\ &= J_e^{-1}(2\boldsymbol{B}_e + \boldsymbol{I}) \cdot \left( \lambda I_1 \boldsymbol{I} + 2G\boldsymbol{B}_e + (lI_1^2 - 2mI_2)\boldsymbol{I} + n\frac{\partial I_3}{\partial \boldsymbol{B}_e} + 2mI_1\boldsymbol{B}_e \right), \end{aligned} \tag{S8}$$

where $J_e = det\boldsymbol{F}_e$ denotes the Jacobian determinant of $\boldsymbol{F}_e$. Compact expressions for $\frac{\partial I_1}{\partial \boldsymbol{B}_e}$, $\frac{\partial I_2}{\partial \boldsymbol{B}_e}$, and $\frac{\partial I_3}{\partial \boldsymbol{B}_e}$ are:

$$\frac{\partial I_1}{\partial \boldsymbol{B}_e} = \boldsymbol{I}; \quad \frac{\partial I_2}{\partial \boldsymbol{B}_e} = -\boldsymbol{B}_e + I_1\boldsymbol{I}; \quad \frac{\partial I_3}{\partial \boldsymbol{B}_e} = \boldsymbol{B}_e \cdot \boldsymbol{B}_e - I_1\boldsymbol{B}_e + I_2\boldsymbol{I}. \tag{S9}$$

*Yield surface:*

$$\phi = \sqrt{3/2\boldsymbol{s} : \boldsymbol{s}} - (\sigma_{y0} + bp) = 0; \sigma_{y0} = (1-c)\sigma_{y0}^{\alpha} + c\sigma_{y0}^{\omega}; b = (1-c)b_1 + cb_2. \tag{S10}$$

Here, $\boldsymbol{s}$ represents the deviatoric part of the Cauchy stress $\boldsymbol{\sigma}$; $p$ denotes the pressure; $\sigma^{\alpha}_{y0}$ and $\sigma^{\omega}_{y0}$ are the yield strengths in compression of the $\alpha$- and $\omega$-Zr @$p = 0$, respectively; $b_1$ and $b_2$ are their respective linear pressure-hardening coefficients.

*Plastic flow rule:*

$$\boldsymbol{\gamma} = |\boldsymbol{\gamma}| \frac{\boldsymbol{s}}{\sqrt{\boldsymbol{s} : \boldsymbol{s}}} = |\boldsymbol{\gamma}|\boldsymbol{n}; \qquad |\boldsymbol{\gamma}| = (\boldsymbol{\gamma} : \boldsymbol{\gamma})^{0.5}. \tag{S11}$$

In the elastoplastic region ($\phi(\boldsymbol{s}, p, c) = 0$), $|\boldsymbol{\gamma}|$ is determined from the consistency condition $\dot{\phi}(\boldsymbol{s}, p, c) = 0$; whereas $|\boldsymbol{\gamma}| = 0$ in the elastic region ($\phi(\boldsymbol{s}, p, c) < 0$ or $\phi(\boldsymbol{s}, p, c) = 0$ and $\dot{\phi}(\boldsymbol{s}, p, c) < 0$).

*Accumulated plastic strain:*

$$\dot{q} = \sqrt{2/3}|\boldsymbol{\gamma}|. \tag{S12}$$

*Pressure-dependence of the yield strengths of $\alpha$- and $\omega$-Zr:*

$$\sigma^{\alpha}_{y} = 0.82 + 0.19p \text{ (GPa)} \quad \text{and} \quad \sigma^{\omega}_{y} = 1.66 + 0.083p \text{ (GPa)}.$$

They were estimated using the peak broadening method [3] near the center of a sample [1].

*Plastic-strain induced kinetic equation:*

$$\frac{dc}{dq} = k(1 + \delta_1 q)(1 + \delta_3 p)(1 - \delta_2 c) \frac{\sigma^{\omega}_{y}}{c\sigma^{\alpha}_{y0} + (1 - c)\sigma^{\omega}_{y0}} \left( \frac{p(q) - p^d_{\varepsilon}}{p^d_h - p^d_{\varepsilon}} \right) H(p - p^d_{\varepsilon}) = A(p, q, c); \tag{S13}$$

$$p^d_{\varepsilon} = \delta_4 + \delta_5 q_0.$$

Here, $p^d_h$ is the pressure for initiation of pressure-induced PT under hydrostatic loading; $p^d_{\varepsilon}$ is the minimum pressure for initiation of the plastic strain-induced PT; $p(q)$ is the loading path; $\sigma^{\alpha}_{y0}$ and $\sigma^{\omega}_{y0}$ are the yield strengths of the $\alpha$ and $\omega$ phases under ambient pressure, respectively; $q_0$ is the value of $q$ at the beginning of PT; $k$ is the kinetic coefficient; $\delta_1$, $\delta_2$, $\delta_3$, $\delta_4$ and $\delta_5$ are the material parameters; $H(x)$ is the Heaviside function, such that $H(x) = 1$ for $x > 0$ & $H(x) = 0$ for $x \leq 0$. The parameters $\delta_1$, $\delta_2$, $\delta_3$, $\delta_4$ and $\delta_5$ are obtained using minimization of cumulative error $Er$ in $\bar{c}(r)$ for all loadings and radii between FEM and experiment:

$$Er(k, \delta_1, \delta_2, \delta_3, \delta_4, \delta_5) = \sum_{\text{all load cases}} \sqrt{\sum_{\text{all radial points}} (\bar{c}_{exp} - \bar{c}_{FEM})^2} \tag{S14}$$

An alternative error indicator is the magnitude of the difference $\bar{\Delta} c_{max}$ between experimental and FEM results, maximized over the radial points and loads.

*Equilibrium equation:*

$$\nabla \cdot \boldsymbol{\sigma} = \mathbf{0}. \tag{S15}$$

## 2.2 Geometry and boundary conditions

Fig. 1a in the main text illustrates the geometry of the DAC. An axisymmetric problem formulation is considered. The geometry of the sample and the anvil, along with the boundary conditions, are the same as in [1]. Simulations employ quadrilateral 4-node bilinear axisymmetric finite elements (CGAX4R), commonly used for large-deformation axisymmetric problems [4]. Our simulations utilize a mesh with 3958 elements.

**Friction model:**

At the culet portion of the diamond $r \leq r_c$, the contact shear stress is given by

$$\tau_{con} = m(r)\tau_y(p) \qquad \text{for } r \leq r_c, \tag{S16}$$

where the distribution of $m(r)$ is obtained from the CEA approach described in [1].

At the inclined portion of the sample-diamond contact surface, the critical shear stress is governed by the combined Coulomb friction $\tau_{cr} = \mu(\sigma_{con})\sigma_{con}$ and $m(r_c)\tau_y(p)$, where $\sigma_{con}$ is the contact normal stress. There is complete cohesion between the sample and the anvil unless the shear stress $\tau_{con}$ reaches the critical value $\tau_{cr}$, i.e.,

$$\tau_{con} < \tau_{cr} = min\left[\mu(\sigma_{con})\sigma_{con}, m(r_c)\tau_y(p)\right] \rightarrow cohesion, \tag{S17}$$

and when the friction stress reaches $\tau_{cr}$, contact sliding occurs, i.e.,

$$\tau_{con} = \tau_{cr} = min[\mu(\sigma_{con})\sigma_{con}, m(r_c)\tau_y(p)] \rightarrow sliding. \tag{S18}$$

## 2.3 Nonlinear elastic equations and material properties for single-crystal diamond anvil

The constitutive response of diamond is modeled using fourth-order nonlinear anisotropic elastic potential energy $\psi(\boldsymbol{E}_e)$ given in [1].
Based on the elasticity law, the Cauchy stress in the diamond can be obtained using:

$$\boldsymbol{\sigma} = \frac{1}{J_e}\boldsymbol{F}_e \cdot \frac{\partial \psi}{\partial \boldsymbol{E}_e} \cdot \boldsymbol{F}_e^t. \tag{S19}$$

Here, $\boldsymbol{F}_e$ is the elastic deformation gradient, and $J_e = det(F_e)$ is its Jacobian determinant.

All the elastic constants of diamond are from [1] and explicitly given below (all in GPa):

$$\begin{aligned} &c_{11} = 1081.9, c_{12} = 125.2, c_{44} = 578.6; \\ &c_{111} = -7611, c_{112} = -1637, c_{123} = 604, c_{144} = -199, c_{166} = -2799, c_{155} = -2799, \\ &c_{456} = -1148, c_{1111} = 26687, c_{1112} = 9459, c_{1122} = 6074, c_{1123} = -425, c_{1144} = -1385, \\ &c_{1155} = 10741, c_{1255} = -264, c_{1266} = 8192, c_{1456} = 487, c_{4444} = 11328, c_{4455} = 528. \end{aligned} \tag{S20}$$

### 2.4 Elastic properties of polycrystalline $\alpha$- and $\omega$-Zr

The elastic constitutive response of polycrystalline Zr is modeled using the third-order nonlinear Murnaghan potential given in Eq. (S5). Among the 5 elastic constants in the Murnaghan potential, Lame constant $\lambda$ and shear modulus $G$ are associated with the quadratic terms in $\boldsymbol{B}_e$, while the remaining constants $l$, $m$ and $n$, are associated with the cubic terms in $\boldsymbol{B}_e$. These constants are calibrated using the bulk modulus $K$ and its pressure derivative $\frac{dK}{dp}@p = 0$, and the shear modulus $G$ and its pressure-derivative $\frac{dG}{dp}@p = 0$. $K@p = 0$ and $\frac{dK}{dp}@p = 0$ are obtained from fitting pressure-volume data, obtained in hydrostatic DAC experiments, into the 3rd order Birch-Murnaghan equation of state (EOS). $G@p = 0$ and $\frac{dG}{dp}@p = 0$ are taken from the experimental results [5,6]. The expressions relating the Murnaghan constants $\lambda$, $G$, $l$, $m$, $n$ and $K$, $G$, $\frac{dK}{dp}$ and $\frac{dG}{dp}$ @ $p = 0$ are as follows:

$$K = \frac{3\lambda + 2G}{3}; \quad \frac{dK}{dp} = K^{'} = -\frac{2(9l + n)}{9K}; \quad \frac{dG}{dp} = G^{'} = \frac{-2G - 6K - 6m + n}{6K}. \tag{S21}$$

It can be observed that there are only 2 equations to solve for the 3 third-order constants. Therefore, there is an indeterminacy of degree 1. However, it can be easily demonstrated that for any pressure, when the deviatoric part of the superposed deformation is small, the stresses and energy can be completely expressed in terms of $K$, $G$, $K'$, and $G'$ only. Hence, one of the constants, $l$, $m$, or $n$, can be arbitrarily chosen, and the other two are determined from Eq. (S21). The constants used are (all in GPa):

$$\begin{aligned} &\lambda = 68.11, \quad G = 36.13, \quad l = -147.01, m = -122.75, \quad n = -100 \qquad \text{for } \alpha - Zr; \\ &\lambda = 72.33, \quad G = 45.1, \quad l = -149.56, m = -179.53, \quad n = -4 \qquad \text{for } \omega - Zr. \end{aligned} \tag{S22}$$

## 3 Supplementary figures

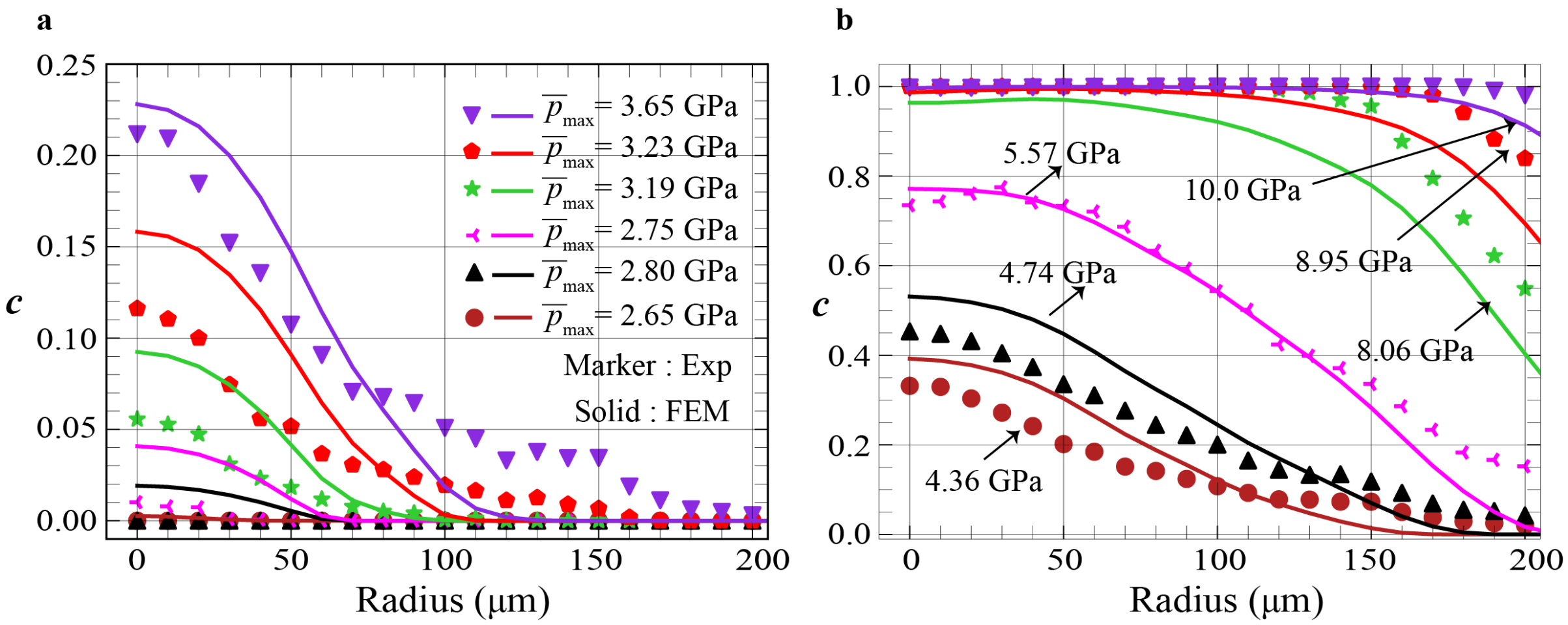

Supplementary Fig. S2: **Comparison of the simulated and experimental radial distributions** *of* $\bar{c}$ **for kinetic equation formulated in [7] and calibrated in [1] with** $\delta_1 = \delta_3 = \delta_5 = 0$**,** $\delta_2 = 1$**,** $k = 5.87$**, and** $p_\varepsilon^d = \delta_4 = 2.65$ **GPa. a Comparison of** $\bar{c}$ **for the lower 6 loadings marked by the averaged pressure at the symmetry axis. b Comparison of** $\bar{c}$ **for the higher 6 loadings.**

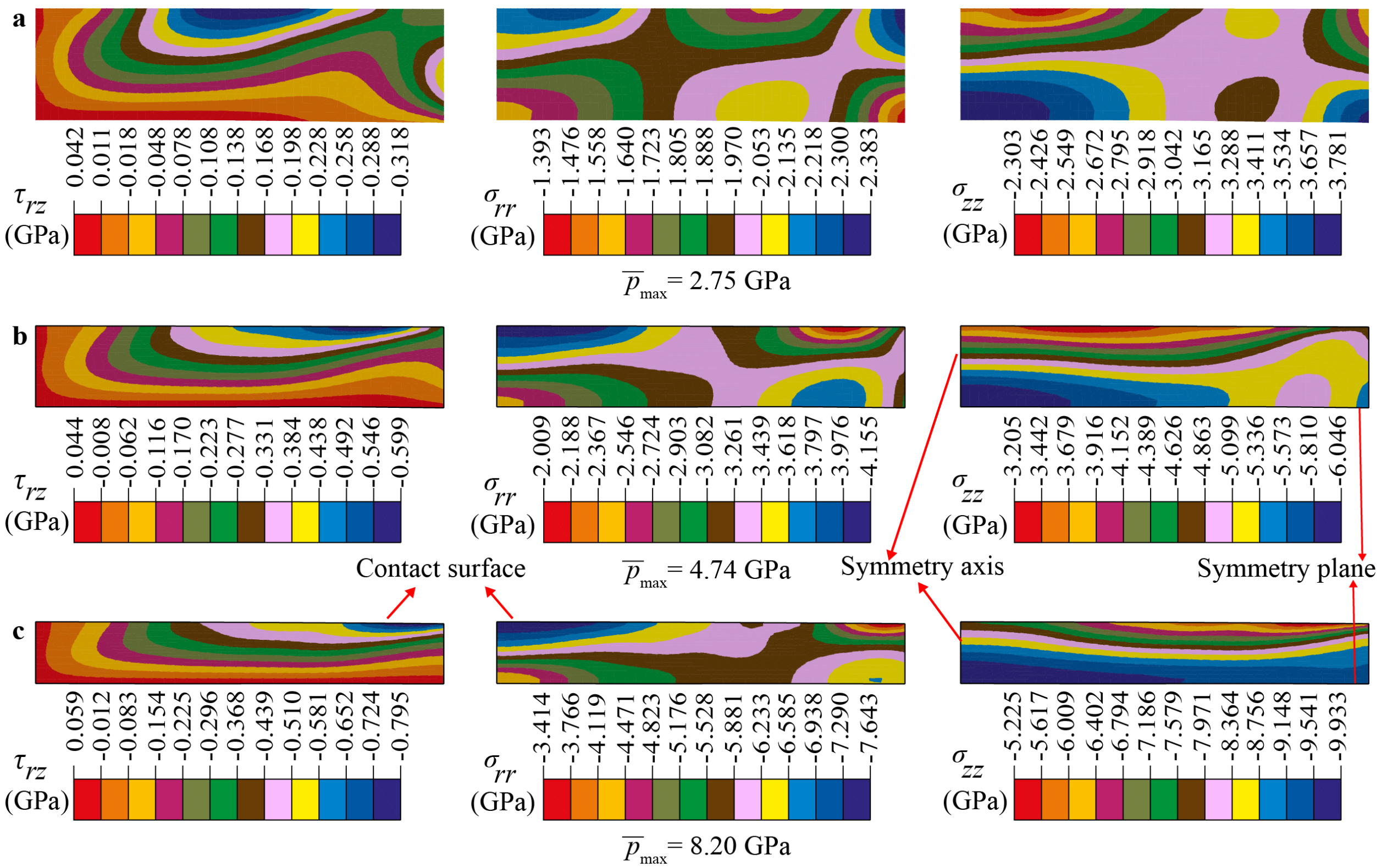

Supplementary Fig. S3: **Distributions of components of Cauchy stress in a sample with parameters $\delta_1 = \delta_3 = 0$, $\delta_2 = 0.803$, $k = 5.20$, and $p_\varepsilon^d = 2.65 - (q_0 - 0.42)$ in the kinetic equation for three loadings.** **a** Results for almost pure $\alpha$-Zr at $\bar{p}_{\max} = 2.75$ GPa. **b** Results for a mixture of $\alpha$- and $\omega$-Zr at $\bar{p}_{\max} = 4.74$ GPa. **c** Results for a mixture of $\alpha$- and $\omega$-Zr at $\bar{p}_{\max} = 8.20$ GPa. The symmetry plane, symmetric axis, and contact surface are marked on the contours.

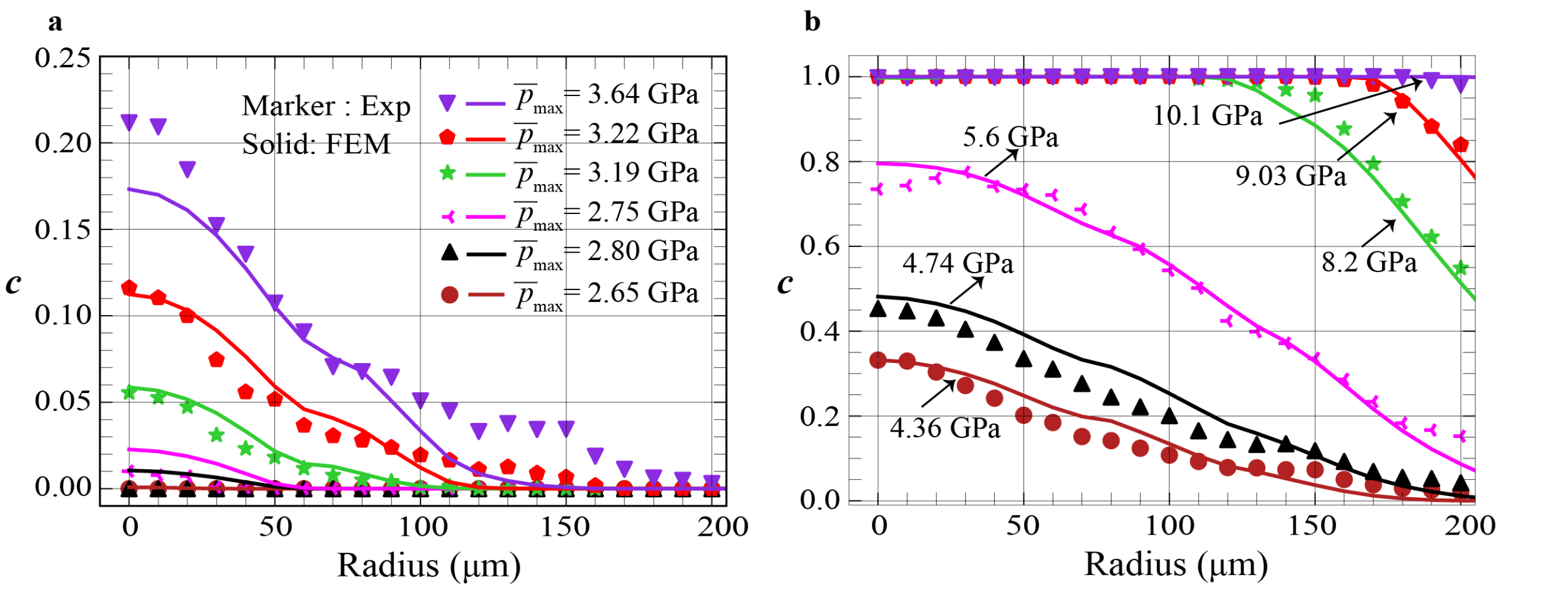

Supplementary Fig. S4: **Comparison of the simulated and experimental radial distributions** *of* $\bar{c}$ **for the model with $\delta_1 = \delta_3 = 0$, $\delta_2 = 0.775$, $k = 5.0$, and $p_\varepsilon^d = 2.65 - (q_0 - 0.42)$ and minimized error $Er = 0.869$.**

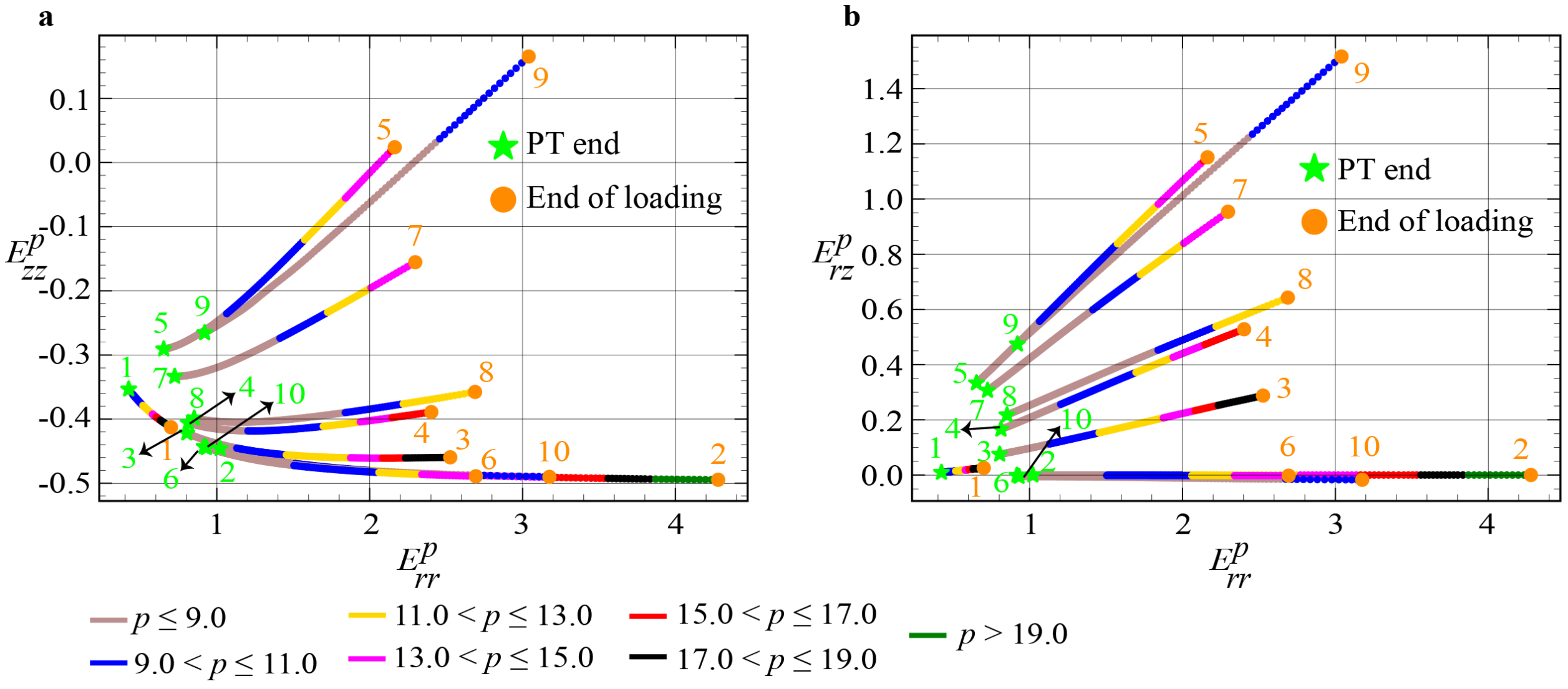

Supplementary Fig. S5: **2D projections of the 4D plastic strain–pressure trajectories from the end of PT to the end of loading. a** and **b** Shear $E_{p,rz}$ – radial $E_{p,rr}$ plastic strains and axial $E_{p,zz}$ – radial $E_{p,rr}$ plastic strains trajectories for various material points with superposed pressure $p$ evolution (in GPa) with parameters $\delta_1 = \delta_3 = 0$, $\delta_2 = 0.803$, $k = 5.20$, and $p^d_\varepsilon = 2.65 - (q_0 - 0.42)$ in the kinetic equation.

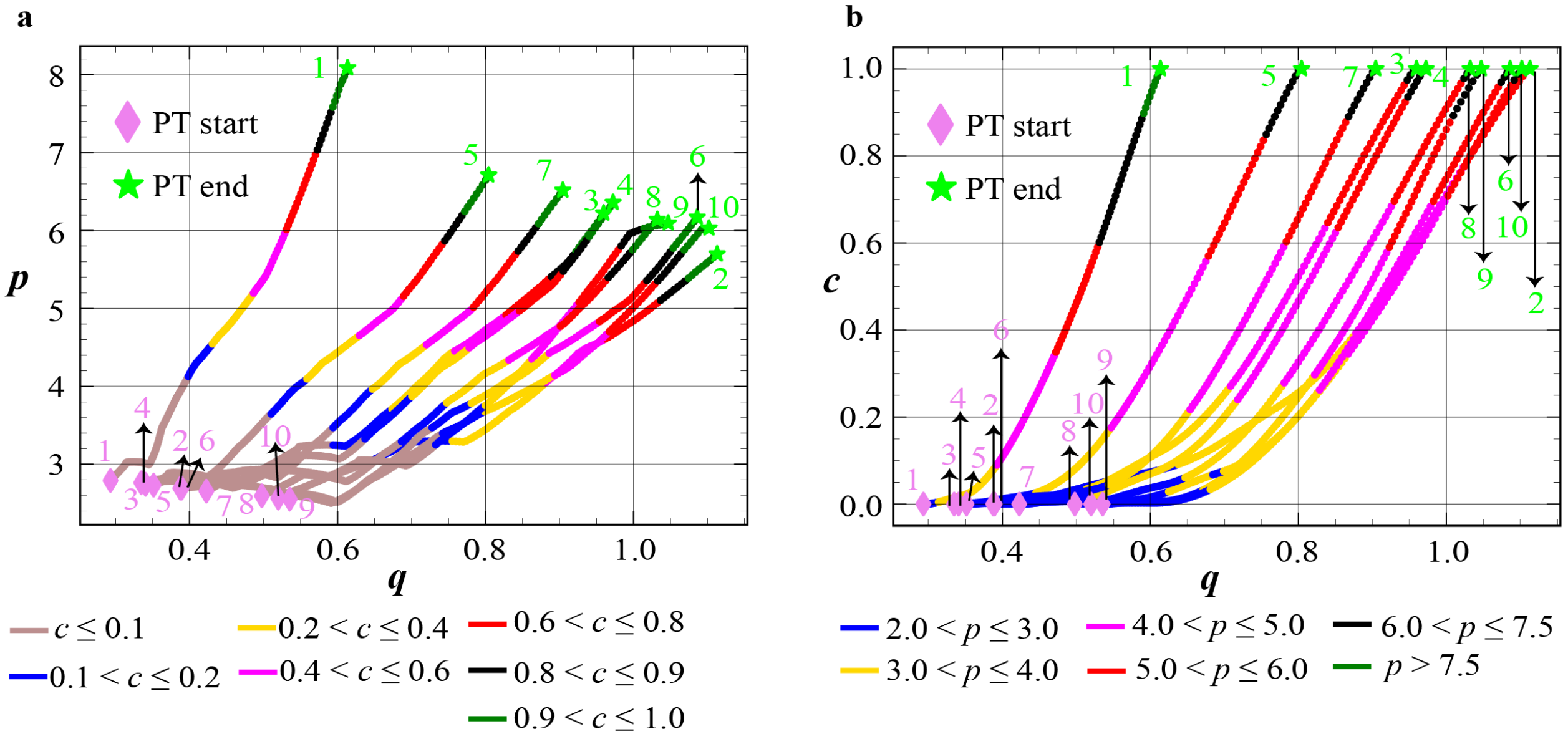

Supplementary Fig. S6: **a** Pressure-accumulated plastic strain-volume fraction loading path trajectories from the start of PT to the end of PT for various material points. **b** Evolution of volume fraction $c$ of $\omega$-Zr as a function of accumulated plastic strain superposed by the pressure (in GPa) for various material points. The parameters used in the kinetic equation are $\delta_1 = \delta_3 = 0$, $\delta_2 = 0.803$, $k = 5.20$, and $p^d_\varepsilon = 2.65 - (q_0 - 0.42)$.

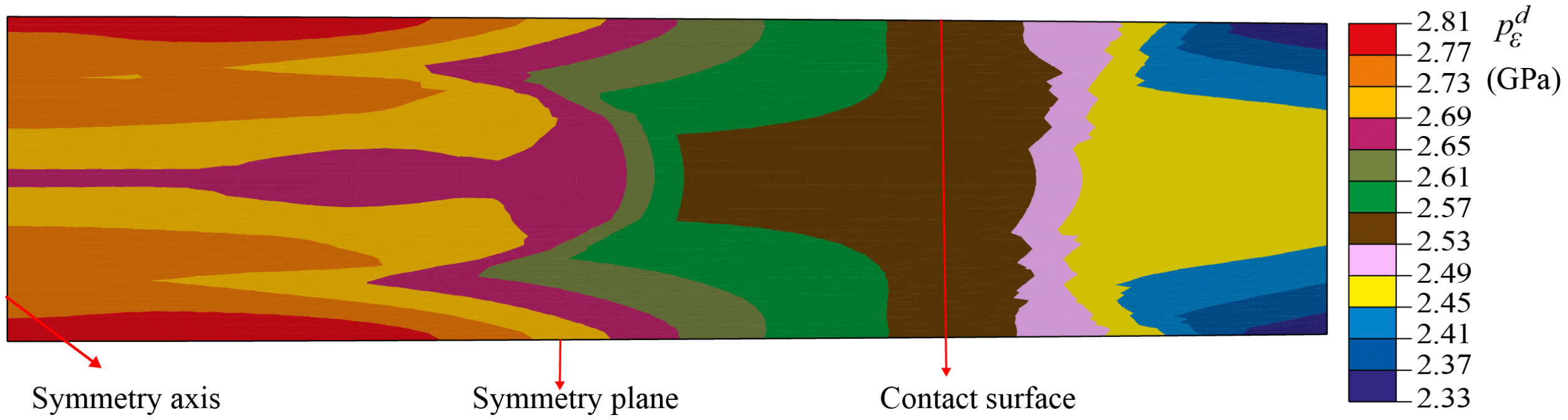

Supplementary Fig. S7: **Distribution of $p_\varepsilon^d$ for the case with parameters $\delta_1 = \delta_3 = 0$, $\delta_2 = 0.803$, $k = 5.20$, and $p_\varepsilon^d = 2.65 - (q_0 - 0.42)$ in the kinetic equation.**

## Supplementary References

[1] Levitas, V. I., Dhar, A. & Pandey, K. K. Tensorial stress-plastic strain fields in $\alpha$- $\omega$ Zr mixture, transformation kinetics, and friction in diamond anvil cell. *Nature Communication*, **14**, 5955 (2023).

[2] Feng, B. & Levitas, V. I. Coupled elastoplasticity and plastic strain-induced phase transformation under high pressure and large strains: Formulation and application to BN sample compressed in a diamond anvil cell. *Int. J. Plast.* **96**, 156-181 (2017).

[3] Zhao, Y. & J. Zhang, Enhancement of yield strength in zirconium metal through high-pressure induced structural phase transition. *Applied Physics Letters*, **91**, 201907 (2007).

[4] Dunne, F. & Petrinic, N. Introduction to computational plasticity (OUP Oxford, 2005).

[5] Liu, W., Li, B., Wang, L., Zhang, J. & Zhao, Y. Elasticity of $\omega$-phase zirconium. *Physical Review B*, **76**, 144107 (2007).

[6] Liu, W., Li, B., Wang, L., Zhang, J. & Zhao, Y. Simultaneous ultrasonic and synchrotron x-ray studies on pressure induced $\alpha$-$\omega$ phase transition in zirconium. *Journal of Applied Physics*, **104**, 076102 (2008).

[7] V. I. Levitas, High-Pressure Mechanochemistry: Conceptual Multiscale Theory and Interpretation of Experiments, Phys. Rev. B **70**, 184118 (2004).